\documentclass[11pt,a4paper]{article}
\usepackage[T1]{fontenc}
\usepackage[utf8]{inputenc}
\usepackage{authblk}
\usepackage{graphicx}
\usepackage[a4paper, total={6in, 9in}]{geometry}
\usepackage{float}
\usepackage{caption}
\usepackage{hyperref}
\usepackage{subcaption}
\usepackage{lmodern,babel,adjustbox,booktabs,multirow}
\usepackage{natbib}
\usepackage{url}
\usepackage{xcolor}
\usepackage{color, colortbl}
\definecolor{Gray}{gray}{0.9}
\usepackage{amsmath, amssymb}
\newtheorem{definition}{Definition}

\newtheorem{theorem}{Theorem}
\newtheorem{proof}{Proof}
\newcommand*{\angela}[1]{\textcolor{black}{#1}}
\newcommand*{\angelaa}[1]{\textcolor{black}{#1}}

\date{}

\begin{document}

\title{Permutation-based true discovery proportions for functional Magnetic Resonance Imaging cluster analysis} 

\author[1]{Angela Andreella*}

\author[2]{Jesse Hemerik}

\author[3]{Livio Finos}

\author[4]{Wouter Weeda}

\author[5]{Jelle Goeman}
\affil[1]{\small Department of Economics, University Ca' Foscari Venezia, Venezia, Italy}
\affil[2]{Biometris, Wageningen University and Research, Wageningen, The Netherlands}
\affil[3]{Department of Developmental Psychology and Socialization, University of Padova, Padova, Italy}
\affil[4]{Department of Psychology, Leiden University, Leiden, The Netherlands}
\affil[5]{Department of Biomedical Data Sciences, Leiden University Medical Center, Leiden, The Netherlands}
\affil[*]{\small correspondig author: angela.andreella@unive.it}

\maketitle

\begin{abstract}
We propose a permutation-based method for testing a large collection of hypotheses simultaneously. Our method provides lower bounds for the number of true discoveries in any selected subset of hypotheses. These bounds are simultaneously valid with high confidence. \angela{The methodology} is particularly useful in functional Magnetic Resonance Imaging cluster analysis, where it provides a confidence statement on the percentage of truly activated voxels within clusters of voxels, avoiding the well-known spatial specificity paradox. We offer a user-friendly tool to estimate the percentage of true discoveries for each cluster while controlling the family-wise error rate for multiple testing and taking into account that the cluster was chosen in a data-driven way. The method adapts to the spatial correlation structure that characterizes functional Magnetic Resonance Imaging data, gaining power over parametric approaches.

\end{abstract}

\section{Introduction}
Functional Magnetic Resonance Imaging (fMRI) is the most frequently used technique to understand which regions of the human brain are activated as a consequence of a stimulus. Brain activation is measured as the correlation between a sequence of (cognitive) stimuli and the resulting blood oxygenation level dependent (BOLD) signal. The BOLD signal over the entire brain is measured in small cubes termed voxels, and for each of these, we test for significant BOLD activity. Typically, around $300{,}000$ voxels are analyzed, so the resulting multiple testing problem has roughly $300{,}000$ statistical tests. 

Controlling Type I error at the voxel level usually negatively affects the power to detect activation \citep{NicholsRev}. Therefore, cluster-extent based thresholding was developed to analyze the data at the level of clusters of contiguous voxels. This method is less conservative than voxel-wise inference since it exploits the spatial nature of the signal \angela{using Random Field Theory (RFT)} \citep{worsley1996unified}. However, the assumptions behind this method require a very high initial cluster-forming threshold, resulting in relatively small clusters left for significance testing \citep{Eklund}. Moreover, the method suffers from the spatial specificity paradox \citep{Woo} meaning that the larger the cluster we find, the less we can say about the signal within that cluster. Since the method tests the hypothesis that none of the voxels in the cluster are active, rejecting this null hypothesis only allows the claim that there is at least one active voxel inside the cluster, hence, the larger the cluster, the less we can say about it. That is, the number of active voxels and their spatial location remains unknown, and doing follow-up inference inside the cluster (``drilling down'') leads to a ``double-dipping'' problem and inflated Type I error rate \citep{Kriegeskorte}.

These problems motivated \cite{ARI} to propose All-Resolution Inference (ARI), a method to compute the lower confidence bound for the true number of active voxels within a cluster (true discovery proportion (TDP)), simultaneously for all possible sets, e.g.,  all clusters of voxels. Simultaneous control permits users to drill down within clusters while maintaining error guarantees, thus resolving the spatial specificity paradox. ARI is based on the approach proposed by \cite{cherry} using closed testing \citep{Marcus} with local Simes test \citep{Simes} to control the family-wise error rate (FWER). The closed testing method has an exponential computational load in general; nevertheless, for this specific case, \cite{Meijer} and \cite{Meijer2} proposed a fast and exact linear time short-cut. ARI relies on the Simes inequality, assuming positive regression dependency on subsets (PRDS) \citep{Sarkar}. While the Simes inequality can be assumed to be valid for fMRI data \citep{NicholsRev} it can be conservative under strong positive dependence. This makes the method inefficient in the neuroimaging data framework, since brain measurements have strong spatial dependence due to both physics and physiology. 

Permutation tests assume only exchangeability under the null hypothesis \citep{Pesarin} and can handle data having any correlation structure, adapting to that correlation structure both to keep type I error control and to gain power. \cite{Jesse} proposed a permutation-based method, related to ARI, that adapts the procedure to the correlation structure of the $p$-values. However, this method finds TDP only for sets consisting of the smallest $k$ $p$-values, simultaneously over $k$, and can therefore not handle spatially defined clusters. Moreover, the method allows much freedom in the choice of the shape of its rejection curve, and the optimal choice for fMRI data is not clear.

In this paper, we merge the strengths of ARI with the permutation-based method of \cite{Jesse} adapting ARI to use permutations using the approach of \cite{Jesse}. The new method provides a lower bound for the TDP for all brain regions, allowing regions of interest to be chosen post-hoc, as in ARI, without compromising family-wise error control. By using permutation-based test statistics, the method gains in power compared to the parametric version of ARI because it adapts to the correlation structure. Moreover, permutation tests are robust, as widely demonstrated in the neuroimaging literature \citep{Eklund, Winkler, Winkler1} and can be used when the parametric assumptions of ARI are not satisfied. The permutation-based post-hoc method proposed here is similar to the one offered by \cite{Neuvial3} which essentially generalized the approach of \cite{Jesse} to arbitrary subsets of the hypotheses. However, we also propose here an iterative approach based on the idea presented by \cite{Jesse} which uniformly improves \cite{Neuvial3}'s method in most cases.

This paper is organized as follows. Section \ref{ARI} introduces the concept of closed testing based on a critical vector, revisiting the results from \cite{GoemanSolari} and \cite{ARI}. Then, in Section \ref{methods}, we combine these results to obtain a permutation-based ARI, and its iterative version in Section \ref{iterative}. We discuss the families of critical curves to be used in Section \ref{lambdaCal} and which test and permutations we recommend to use in fMRI data in Section \ref{one-sample-location-test}. Section \ref{applications} evaluates the performance of our method in comparison with the parametric version in fMRI data.  We validate the method using the resting-state fMRI null data of \cite{Eklund} in Section \ref{validiting-and-power-of-the-permutation-based-ari-inference}. Finally, we perform some simulations in Section \ref{Simulation} in order to investigate the influence of the shape of the rejection curve in different scenarios. 

\section{Closed testing for true discovery proportions}\label{ARI}

In this section, we revisit some results from \cite{ARI} and \cite{GoemanSolari} to introduce notation and to clarify the need for selective inference in fMRI data. 

Suppose the brain $B$, with $|B|=m$, is composed of $m$ voxels, and let $2^B$ be the collection of all subsets of the brain. Some of the voxels are truly active: let $A \subseteq B$ be the unknown set of all truly active voxels. For a cluster of interest $S \subseteq B$ we want to make inference on $a(S) = | A \cap S |$, i.e., the number of truly active voxels in $S$, or equivalently the TDP, i.e.,  $a(S)/|S|$.  

We assume that we have computed a test statistic for each voxel $i$, where $i = 1, \dots, m$, corresponding to the null hypothesis that the voxel is not active. Based on some knowledge or guess of the marginal null distribution of these test statistics, we may compute the corresponding (parametric) $p$-values $p_i: \Omega \rightarrow [0,1]$ where $\Omega$ is the sample space of the data $X$. In the parametric version, we will assume that these $p$-values will be valid, i.e., stochastically smaller than the uniform distribution of all inactive voxels. For the permutation-based method, we emphasize here that, to guarantee FWER control, we do not make any assumptions on the distribution of these $p$-values. The reason is that it will suffice that the p-values are computed in the same way for all permuted versions of the data \citep{Jesse}.

We will now revisit the simultaneous inference on TDP using closed testing and critical vectors. First, we define a critical vector.
\begin{definition}\label{cv}
	A vector $(l_1,\ldots, l_m)$ is a critical vector if and only if
	\begin{align}\label{eq3}
		\Pr(\cap_{i=1}^{|N|}\{q_{(i)} \ge l_{i}\}) \ge 1-\alpha,
	\end{align}
	where $N = B \setminus A$ is the set of inactive voxels, and $q_{(i)}$,  $1 \le i \le |N|$, are their sorted $p$-values. 
\end{definition}

The general parametric version of ARI assumes that, for a chosen error rate $\alpha \in [0,1]$, there is a critical vector $(l_1,\ldots, l_m)$, possibly random, expressed as Definition \ref{cv}. If such a critical vector exists, then as a corollary to Lemma $6$ from \cite{GoemanSolari} we have the following proposition, which we prove in Appendix \ref{thm_proof}. 

\begin{theorem}\label{prop}
	Let $l_i$ satisfy \eqref{eq3}. Then for every $\emptyset \ne S \subseteq B$,
	\begin{align}\label{as}
		\bar{a}(S) = \max_{1 \le u \le |S|} 1 - u + |\{i \in S: p_i \le l_u \}|
	\end{align}
	is a lower $(1-\alpha)$ confidence bound of $a(S)$, simultaneously for all $ S \subseteq B$, that is
	\begin{align}\label{FWER}
		\Pr(\forall S \subseteq B: \bar{a}(S) \le a(S)) \ge 1-\alpha.
	\end{align}
\end{theorem}

In ARI, the Simes-based critical vector is $l_i = i\alpha/h$, where $h$ is a random variable that can be calculated using the short-cut defined by \cite{Meijer}. It is the largest set size of a subset of the brain not rejected by the Simes test. 

\angela{
	The multiplicity control \eqref{FWER} that ARI guarantees is very versatile. It guarantees,  simultaneously for every subset $S$ of the brain, that the true activation $a(S)$ is at least as large as the claimed activation $\bar{a}(S)$. The analogous result for TDP follows immediately. Several more familiar error rates can be derived from \eqref{FWER}. Taking all clusters with TDP$=1$ is equivalent to strong control of FWER at the voxel level. Taking clusters with TDP$>0$ is equivalent to strong control of FWER at the cluster level but weak FWER control at the voxel level. At intermediate levels of TDP, \eqref{FWER} gives intermediate information between weak and strong control at the voxel-level. For more about relationships between \eqref{FWER} and classical error rates, see \cite{GoemanSolari}. Note that standard cluster-wise approaches based on the RFT or permutations \citep{nicholsprimer,smith2009threshold} only provide strong control of the FWER at the cluster-level and weak control at the voxel-level, which is one of the error rates implied by \eqref{FWER}.
	The second feature, perhaps even more relevant, of \eqref{FWER} is that the inference is simultaneous over all possible subsets of tested hypotheses (i.e., voxels). Simultaneously implies that any exploratory and iterative approaches (i.e., double-dipping) that are not possible in the cluster-wise approach become valid in the ARI class of methods. That is, the inferences on all subsets $S$ are valid simultaneously and regardless of how they were selected (after seeing the data, changing the cluster-wise threshold, etc.).}

\begin{figure}[!ht]
	\centering
	\includegraphics[width=.8\textwidth]{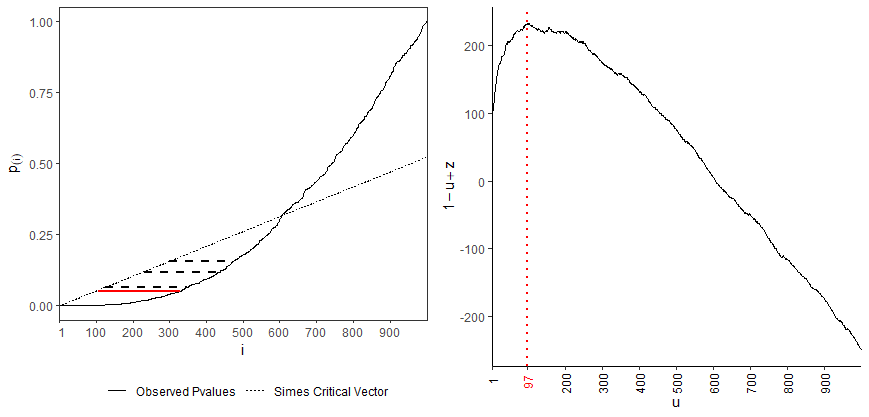}
	\caption{Left Figure: A graphical display of the computation of $\bar{a}(S)$. The sorted p-values $p_{(i)}$ are plotted as solid line against their indexes $i$ for $i = 1,\dots, 1000$. $\bar{a}(S)$ equals then the highest horizontal distance (solid red segment) between the curve of observed $p$-values $p_{(i)}$ and critical vector (dotted line) considering the distances represented by the black dashed long lines (here we put a sample). Right Figure: Example of the computation of $\bar{a}(S)$, i.e., the maximum $1-u+z$ over $u$ where $u \in \{1, \dots, 1000\}$ and $z = 1 - u + |\{i \in S: p_i \le l_{u} \}|$. $\bar{a}(S)$ is then the maximum value of $1-u + z$, represented by the red dotted line, attained when $u$ equals $97$.}
	\label{fig:example}
\end{figure}

Figure \ref{fig:example} illustrates computation of $\bar{a}(S)$ as defined in Equation \eqref{as}, where $|S|=1000$. In the left part, the length of the dashed black segments are the $z=1 - u + |\{i \in S: p_i \le l_{u} \}|$ with $u \in \{1, \dots, |S| \}$ described in Equation \eqref{as}, while the solid red segment is the maximum value over $u$, i.e., the highest distance between the curve of observed p-values and critical vector $l_{i}$, e.g., Simes-based. In the right part, we can see the trend of $\bar{a}(S)$ over $u$. The maximum value of $1-u + z = 232$ is reached when $u$ equals $97$. This implies that $\bar a(S)= \max_{1 \le u \le |S|} 1 - u + |\{i \in S: p_i \le l_u \}| = 232$ is a lower confidence bound for the number of true discoveries in $S$.

The crucial assumption of Theorem \ref{prop} is that $l_i$ satisfies \angela{Equation} \eqref{eq3}. In the case of the Simes test used by ARI, this follows from the PRDS assumption, commonly also adopted for the False Discovery Rate (FDR) controlling approach proposed by \cite{Benjamin}. Although this assumption is commonly accepted in neuroimaging \citep{NicholsRev}, the critical values $(l_1,\ldots, l_m)$ can be overly strict if $p$-values are positively correlated, leading to conservative results. Moreover, the Simes critical vector may also be too strict or too loose if the $p$-values are not well calibrated.

\section{Permutation-based All-Resolutions Inference}\label{methods}
To obtain a critical vector that leads to improved power, we propose a permutation procedure, based on results in \cite{Jesse}. The permutation method takes into account the dependence structure of the $p$-values and, therefore, often leads to a higher critical curve than parametric methods.
Moreover, permutation methods not only adapt to the dependence structure but also to the marginal distributions of the $p$-values. This means that we do not require the null $p$-values to be uniformly distributed. Instead, we require that the null $p$-values are exchangeable with the corresponding post-permutation $p$-values (Assumption $1$ in \cite{Jesse}).

Following \cite{Jesse}, we consider a group of permutations or sign-flipping transformations or any other data transformation that preserves the distribution of the test statistics under the null hypothesis, such as rotations \citep{Hemerik3}. These are maps from the support of the data distribution to itself. Our method is based on $w$ random permutations or sign-flipping transformations. Let $p^1_1,...,p^1_m = p_1,...,p_m$ be the $p$-values for the real data, and for every $2\leq j \leq w$, let $p^j_1,...,p^j_m$ be the $p$-values obtained for the $j$-th random permutation of the data. 

Computing all possible permutations could be computationally infeasible, especially in the fMRI framework. However, Proposition $2$ of \cite{Exact} states that if the permutation set has a group structure, and $\alpha \in [0/w,1/w, \dots (w-1)/w]$, the random permutations reach an exact $\alpha$ level. This means that the $\alpha$ level is exhausted if all hypotheses are true, and the error rate is at most $\alpha$ otherwise. 

To obtain the permutation-based critical vector, the user must choose a family of candidate critical vectors. Examples of such candidate vectors are given in Section \ref{lambdaCal}. We suppose that the candidate vectors are indexed by $\lambda_{\alpha}\in\Lambda\subseteq\mathbb{R}$, so that $l(\lambda_{\alpha})$ denotes the candidate vector corresponding to $\lambda_{\alpha}$. 
The family of candidate vectors is thus $\mathcal{F}=\{l(\lambda_{\alpha}):\lambda_{\alpha}\in\Lambda\}$.
We assume that the family of candidate vectors is monotone, in the sense that if $\lambda_{\alpha}^1,\lambda_{\alpha}^2\in \Lambda$ and $\lambda_{\alpha}^1\leq \lambda_{\alpha}^2$, then $l_i(\lambda_{\alpha}^1)\leq l_i(\lambda_{\alpha}^2)$ for every $1\leq i \leq m$.

We define the permutation-based critical vector to be $l(\lambda_{\alpha})$, where 
\begin{align} \label{lambdaalpha}
	\lambda_{\alpha}=\sup\{\lambda \in \Lambda:  w^{-1}|\{1\leq j \leq w: p^j_i\geq l_i(\lambda) \,\,\, \forall i \in B \} |\geq 1-\alpha\}.
\end{align}
By \cite{Jesse}, the following holds, so that Theorem \ref{prop} applies.

\begin{theorem}\label{thm_cv}
	The vector $l(\lambda_{\alpha})$ is a critical vector, i.e., it satisfies \eqref{eq3} of Definition \ref{cv}.
\end{theorem}

\begin{figure}[!ht]
	\centering
	\includegraphics[width=8cm,height=8cm]{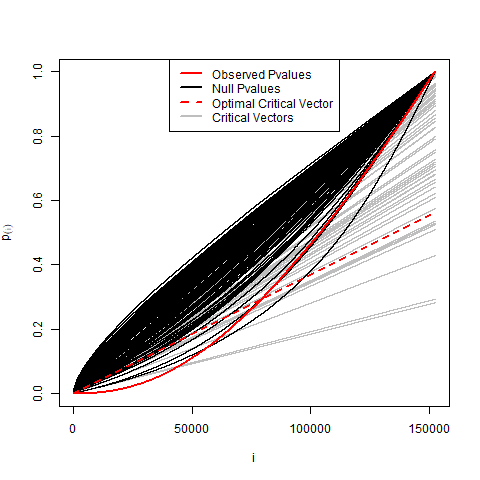}
	\caption{Example of $l_i(\lambda_{\alpha})$ computation using $\alpha = 0.10$. The sorted p-values $p_{(i)}$ are plotted against their indexes $i = 1, \dots, m$ with $m = 150,000$. The dashed red line represents the highest critical curve, i.e., the optimal critical vector $(l_1(\lambda_{\alpha}), \dots, l_m(\lambda_{\alpha}))$ than the gray ones, such that the $\alpha\%$ $p$-values distribution (black curves plus red one) is below it. }
	\label{fig:lambdacal}
\end{figure}

The $\lambda_{\alpha}$-calibration permits to incorporate the unknown dependence structure of the data into the choice of the critical vector. As seen from \angela{Equation} \eqref{lambdaalpha},  the vector $\lambda_{\alpha}$  is computed in such a way that for at least $(1-\alpha)100\%$ of the permutations, all $p$-values lie above it.
This is illustrated in Figure \ref{fig:lambdacal}, considering a random sample of $100$ permutation curves. The $\lambda_{\alpha}$ parameter tends to be lower, if many null hypotheses are false, being "optimal" if Equation \eqref{lambdaalpha} considers only $p_i \in B \setminus A$, i.e., the set of true null hypotheses. 

We use the critical vector $l_{i}(\lambda_{\alpha}) \in \mathcal{F}$ in Theorem \ref{prop} instead of the Simes-based one employed in ARI to gain power in computing $\bar{a}(S)$. \angela{The time complexity to compute the lower confidence bounds for the TDP is $\mathcal{O}(|S| log(|S|))$, after an initial step of calculating the critical values, which takes $\mathcal{O}(wm log(m))$, using a similar algorithm as was used by \cite{Meijer2} for the version with the parametric Simes test. So, it remains close to linear also if the whole brain is analyzed.} Finally, the method can be uniformly improved by its iterative version \citep{Jesse}, presented in the next section.

\section{Iterative approach}\label{iterative}

We propose here an iterative method which uniformly improves $\bar{a}(S)$ defined in Equation \eqref{as} following the idea proposed by \cite{Jesse}. The confidence envelope is defined as the minimum confidence bound computed in the complementary set of the rejection set having a cardinality equal to the lower bound of the number of true discoveries found in the previous iteration. The improvement is then substantial only when the number of detectable false hypotheses is large. In short, the iterative method improves $\bar{a}(S)$ sequentially, in each step using the bound obtained in the previous step. The method always converges after a finite (and usually small) number of steps.

We rephrase below Theorem $2$ of \cite{Jesse} to get an improvement of the calibration parameter $\lambda_\alpha$.

\begin{theorem}\label{iterative_thm1}
	Let $\lambda_{\alpha}^0$ = $\lambda_{\alpha}$ as defined in \eqref{lambdaalpha}, we define $\lambda_{\alpha}(K)$ as,
	\begin{align*} 
		\lambda_{\alpha}(K) =\sup\{\lambda \in \Lambda:  w^{-1}|\{1\leq j \leq w: p^j_i\geq l_i(\lambda) \,\,\, \forall i \in K \} |\geq 1-\alpha\}
	\end{align*}
	where $K^c$ is the complement of $K$. For $i \in \mathbb{N}$ and fixed $c \in [0,1]$ we consider $R = \{x \in B : p_x \le c \} $, and we determine:
	\begin{align*}
		\lambda_{\alpha}^{i + 1} = \min \{\lambda_{\alpha}(K^c): K \in R, |K| = \max_{1 \le u \le |R|} 1 - u + |\{ i \in R : p_i \le l_u (\lambda_{\alpha}^i) \}| \}.
	\end{align*}
	Then $\lambda_{\alpha}^{0} \le \lambda_{\alpha}^{1} \le \cdots$, and for a certain $i \in \mathbb{N}$, $\lambda_{\alpha}^{i} = \lambda_{\alpha}^{i + 1}$. The function $l(\lambda_{\alpha}^{it})$ where $\lambda_{\alpha}^{it} = \max_{i \in \mathbb{N}}  \lambda_{\alpha}^{i}$ is a critical vector in the sense of Definition \eqref{cv}.
\end{theorem}

Theorem \eqref{iterative_thm1} returns a confidence bound which uniformly improves the one defined in Theorem \eqref{prop} with $l(\lambda_{\alpha})$ defined in Theorem \eqref{thm_cv} as critical vector. Furthermore, if in every step we compute the improved bound for all $c\in[0,1]$ and take the best one, then we are always better than the method proposed by \cite{Neuvial3} if the same family of curves $\mathcal{F}$ is used. There are multiple versions of the iterative method, and that one (where in each step we find the best $c$) is a uniform improvement of \cite{Neuvial3}'s method.

In addition, we can simply demonstrate the uniform improvement over \cite{Neuvial3}'s method by setting a specific $c \in [0,1]$. First of all, by definition \eqref{lambdaalpha}, if $K_1 \subseteq K_2$, then $\lambda_{\alpha}(K_1) \ge \lambda_{\alpha}(K_2)$. Consequently $\lambda_{\alpha}(K_1)$ returns a greater confidence bound for $a(S)$ than the one calculated with $\lambda_{\alpha}(K_2)$. We can then focus on the size of the set $K$ used to compute $\lambda_{\alpha}(K)$ to analyze the improvement of the iterative method. In the first step, the \cite{Neuvial3} algorithm computes $\lambda_{\alpha}(K_1)$ with $|K_1| = |\{i \in B : p_i \ge l_1(\lambda_{\alpha}(B))\}| = k$. Instead, our iteration approach computes $\lambda_{\alpha}(K_2)$ where $|K_2| = m - \bar{a}(R)$. By definition of $R$, we can consider $c = l_1(\lambda_{\alpha}(B))$.  Therefore, we have $|R| = m - k$ and $\bar{a}(R) \le |R|$ which implies $|K_2| \le m - m + k = k$, so $|K_2| \le |K_1|$. This leads to $\lambda_{\alpha}(K_2) \ge \lambda_{\alpha}(K_1)$, and then we can say that the lower confidence bound proposed in Theorem \ref{iterative_thm1} uniformly improves the one proposed by \cite{Neuvial3}.

The iterative method is uniformly more powerful than the single-step method defined in Section \ref{methods}, and also is uniformly more powerful than the step-down approach presented by \cite{Neuvial3} under certain conditions. However, the power gain has as a cost a high computational time. In fact, the calculation of $\lambda_{\alpha}^{it}$ can be  computationally infeasible if a large number of hypotheses is considered as in the fMRI scenario. The iterative approach must compute the minimum of a set of size $|S|!/(|S| -\bar{a}^i(S))! \bar{a}^i(S)!$. Nevertheless, we suggest to use the approximated approach defined by \cite{Jesse} which can be directly applied to our case. It simply calculates the minimum across sets randomly sampled from $\{K \in S : |K| = \bar{a}^i (S)\}$ for $i \in \mathbb{N}$. The computation time, in this case, equals approximately \angela{$37$} seconds analyzing $2000$ hypotheses, $20$ observations, and $1000$ permutations. Finally, the approximated iterative approach provided valid inference in all simulations in \cite{Jesse}. Please see Appendix \ref{it_plus} for further details.

\section{Choice of family of curves}\label{lambdaCal}
In the previous section, we consider a general family $\mathcal{F}$ of candidate vectors $l(\lambda_{\alpha})$, $\lambda_{\alpha}\in\Lambda$. Here we will discuss several examples of such families, which we considered in the application later in the paper.

The first family $\mathcal{F}$ that we consider is inspired by  Simes' probability inequality \citep{Simes}. The vectors are obtained by multiplying and shifting the Simes' critical vector. We denote the shift by $\delta \in \{0, \dots, m-1\}$. For every such $\delta$, we have a different family, indexed by $\lambda_{\alpha} \in \mathbb{R}$.
The candidate critical vector $l(\lambda_{\alpha})$ is defined by
\begin{align}\label{eq1}
	l_{i}(\lambda_{\alpha}) = \dfrac{(i  - \delta) \lambda_{\alpha}}{m - \delta}.
\end{align}
The shift parameter $\delta$ can be used to determine how sensitive the critical vector $l(\lambda_{\alpha})$ will be to the smallest $p$-values. The parametric Simes-based approach corresponds to $\delta=0$ and $\lambda_\alpha=\alpha$. We gain over that approach only if the $\lambda_{\alpha}$ value is greater than $\alpha$. 

Regarding the choice of $\delta$, note that $l_i(\lambda_{\alpha})\leq 0$ for $i\leq \delta$. As a consequence, we will find $\bar a(S) \leq |S|-\delta$, and $\bar a(S) =0$ for all $S$ with $|S|\leq \delta$. The value of $\delta$ therefore corresponds to the minimum size of a cluster that we are interested in detecting. To compensate, methods with large $\delta$ will often have a steeper slope $\lambda_\alpha$ and consequently have more power for detecting large clusters. In addition, the lower confidence bound $\bar{a}(S)$ computed by the shifted version, i.e., $\delta>0$, can not reach the $100\%$ true discovery proportion, since the maximum equals to $(|S| - \delta)/|S|$. Further details about the shift parameter in computing bounds for the false discovery proportion can be found in \cite{Katsevich}.

The second example that we propose is a family $\mathcal{F}$ of candidate vectors that are derived from the asymptotically optimal rejection curves (AORC) considered in \cite{Finner} to control the FDR in an asymptotic Dirac uniform setting. Again we add a shift parameter $\delta \in \{0, \dots, m\}$ as above, and we have a different family of candidate vectors for each $\delta$. The calibration parameters lie in $\Lambda\subseteq\mathbb{R}$.
The candidate critical vector $l(\lambda_{\alpha})$ is defined by 
\begin{align}\label{eq2}
	l_{i}(\lambda_{\alpha}) = \dfrac{(i-\delta) \lambda_{\alpha}}{(m - \delta) - (i-\delta)(1- \lambda_{\alpha})}.
\end{align}

Our third example is related to the Higher Criticism method proposed by \cite{Donoho}. The candidate vectors are indexed by $\lambda_{\alpha}\in \Lambda\subseteq \mathbb{R}$ and are given by 
\begin{align}\label{hc}
	l_{i}(\lambda_\alpha) = \dfrac{2i + \lambda_{\alpha}^2 - 
		\sqrt{(2i+ \lambda_{\alpha}^2)^2 - 4 i^2  (m + \lambda_{\alpha}^2)/m)}}{2 (m + \lambda_{\alpha}^2)}.
\end{align}
Finally, we, we consider the family of candidate vectors $l(\lambda_{\alpha})$ defined as follows: 
\begin{align}\label{beta}
	l_{i}(\lambda_\alpha) = \inf \{x: \lambda_{\alpha} \le F_{i}(x) \}.
\end{align}
Here $\lambda_{\alpha} \in \Lambda= [0,1]$ and  $F_{i}(X)$ is the cumulative distribution function of the beta distribution $\text{Beta}(i,m+1-i)$. This family was also considered in \cite{Jesse}.

Further examples of candidate critical vectors can be found in \cite{Neuvial1, Neuvial2, Neuvial3} and \cite{Jesse} but we did not consider them here. 
The results obtained with our permutation method will depend on the critical vector $l(\lambda_{\alpha})$ and hence on the choice of the family $\mathcal{F}=\{l(\lambda_{\alpha}):\lambda_{\alpha}\in\Lambda\}$. However, one choice of a family will essentially never lead to a uniform improvement compared to another family, but only to improved TDP bounds for some sets of hypotheses and worse bounds for other sets of hypotheses. Thus, the most appropriate family will depend on which set of hypotheses we are most interested in, e.g., on whether we are interested in large or small clusters.
Section \ref{applications} provides guidelines regarding a good choice of $\mathcal{F}$ for fMRI data.

\section{Choice of permutations and t-statistic}\label{one-sample-location-test}

\angela{In fMRI activation studies} the correlation between the \angela{(convolved)} sequence of cognitive stimuli and \angela{the changes in} BOLD \angela{response} expresses brain activation. The \angela{changes in local hemodynamics affect} the intensity of the magnetic resonance signal, i.e., the voxel intensity. Therefore, the intensity of each voxel becomes the unit of interest. \angela{The differences in intensity, either between different conditions of an experiment or between different groups of participants, are expressed as a statistic value (t, z, or F usually), with an associated p-value. In fMRI intensity values are often} characterized by high spatial correlation\angela{s} and heteroscedasticity across voxels and across subjects due to \angela{the nature of the BOLD signal and external nuisance factors (}e.g., quality of the recording, \angela{respiration, or heartbeat)}. The permutation approach is useful in this situation, where parametric tests fail due to violations of assumptions.

In this section, we review which permutation test is valid and powerful to perform fMRI group analysis consisting of multi-subject studies to explore the differences in BOLD response recorded under two experimental conditions \citep{Holmes, Winkler2, Helwing}. Group fMRI data are widely analyzed using a two-stage summary statistics approach within a mixed model. This approach uses ordinary least squares (OLS) methods \citep{Mumford1}, in particular one- or two-sample t-tests, using within-subject parameter estimates as observations. 

Let the first level within-subject model for each voxel $i \in \{1, \dots, m\}$ and each subject $j \in \{1, \dots, J\}$:
\begin{align*}
	Y_{ij} = X_j \beta_{ij} + \epsilon_{ij}, 
\end{align*}
where $Y_{ij} \in \mathbb{R}^{n}$ is the brain signal of subject $j$ in voxel $i$, $n$ is the total number of time points, $J$ is the total number of subjects, $m$ is the total number of voxels, $X_j \in \mathbb{R}^{n \times p}$ is the design matrix, where $p$ regressors of interest, $\beta_{ij} \in \mathbb{R}^{p}$ is the vector of parameters, and $\epsilon_{ij} \in \mathbb{R}^n$ is the vector of autocorrelated and non-independent error terms. Let $\beta_{1ij}$ be the parameter relative to the first experimental condition, while $\beta_{2ij}$ to the second experimental condition for the subject $j$, we then assume for simplicity $p=2$. We make inference on the contrasts of parameter estimates involving brain activation differences, i.e., $D_{ij} = \hat\beta_{1ij} - \hat\beta_{2ij}$, so:
\begin{align}\label{one-sample-model}
	D_{ij} = \mu_i + \epsilon_{ij}^\star
\end{align}
where $\mu_i$ is the unknown parameter of interest representing the between-subject mean activation in voxel $i$, and $\epsilon_{ij}^\star$ are the error terms $\sim \mathcal{N}(0, \Sigma)$. To make inference on $\mu_i$, the one-sample t-test is performed for each voxel $i$:
\begin{align}\label{eq:onesam}
	T_i = \frac{\hat{\mu_i}}{\sqrt{\hat{\sigma_i}^2/J}}
\end{align}
where $\hat{\mu_i}$ equals $\sum_{j =1}^{J} D_{ij}/J$ and $\hat{\sigma_i}^2$ equals $\sum_{j =1}^{J} (D_{ij} - \hat{\mu_i})^2/(J-1)$. So, we have $m$ statistical tests to analyze, one for each voxel $i$, i.e., $H_{0i}: \mu_i = 0$, that create a statistical brain mapping. 

Nevertheless, we need valid permutations to have a valid permutation testing procedure. It needs a null-invariant transformation of the data, i.e., the joint distribution of the $p$-values under $H_{0i}$ does not change. \cite{Exact} In this case, $H_{0i}: \mu_i = 0$ implies that $( \beta_{1ij},  \beta_{2ij}) \overset{d}{=} ( \beta_{2ij}, \beta_{1ij})$, that is equivalent to $D_{ij}  \overset{d}{=} - D_{ij} $ for each voxel $i$. The compound symmetry is weaker than normality and also allows for heteroskedasticity. It can be justified by subtraction of two sample means with the same (arbitrary) distribution. Therefore, under $H_{0i}$, we can flip the sign at random of each $D_{ij}$ \citep{Winkler2} always taking the identity permutation as the first transformation to have an exact $\alpha$ method \citep{Exact,Pesarin}. 

The same approach can be used in the case of two-sample t-test. Let $G_j=\{1,2\}$ expresses the group label for the $j$-th subject, the null hypothesis is then defined as $H_{0i}: \mu_{1 i} = \mu_{2 i}$. The exchangeability assumption implies $(D_{ij}|G_j = 1) \overset{d}{=} (D_{ij}|G_j = 2)$ for each voxel $i$, we can just shuffle the subject-group labels at random to compute the $p$-values null distribution. Permutation-based tests can be applied in various hypothesis testing's situations, e.g., tests for linear models even in the presence of nuisance effects \citep{Winkler2, Helwing, Solari}, and tests for generalized linear models \citep{Hemerik3}.

\section{fMRI data Application}\label{applications}
In this section, the permutation-based ARI method is evaluated using fMRI data. Two datasets from \url{https://openfmri.org/} are analyzed. Both datasets have the same experimental design, i.e., a block design with two stimuli. Pre-processing and first-level data analysis were performed using FSL \citep{Jenkinson}. Registration to MNI space was done using FLIRT \citep{Jenkinson1, Jenkinson}, motion correction using MCFLIRT \citep{Jenkinson3}, and brain-extraction using BET \citep{Smith}. We applied spatial smoothing using a Gaussian kernel of 6mm FWHM. Finally, we applied a high-pass filter to the time-series data  (Gaussian-weighted least-squares straight-line fitting, with sigma=64.0s). The parameter estimates (copes), i.e., $\mathbf{D}_j \in \mathbb{R}^{m}$, were used as input in the \texttt{pARI} \citep{ARIperm} package developed in \texttt{R} \citep{R}. These parameter estimates are instead downloadable by installing the \texttt{fMRIdata} \texttt{R} package \citep{fMRIdata}.

For all the analyses, the $\alpha$ level is taken as $0.05$ for a two-sided alternative hypothesis. We use $1000$ permutations: 999 random permutations plus the identity. The approximated iterative approach (100 random combinations) presented in Section \ref{iterative} is then applied. The results using the single-step method, i.e., $\lambda_{\alpha}$ computed on the full set of hypotheses, are reported in Appendix \ref{auditory_plus} and \ref{rhyme_plus}.

We chose $\delta$ as $0$, $1$, $9$, and $27$, to account for signal spreading out in clusters with size at least equals $0$, $1$, $9$ and $27$ voxels. The third powers were considered to exploit the three-dimensional structure of the voxels. 

\subsection{Auditory Data}\label{auditory-data}

We analyzed data from $140$ subjects passively listening to vocal (i.e., speech), and non-vocal sounds, collected by \cite{Pernet}, available at \url{https://openneuro.org/datasets/ds000158/versions/1.0.0}. We estimated the statistics map regarding the contrast that describes the difference of neural activation during vocal and non-vocal stimuli for each participant, i.e., $\mathbf{D}_j$. The hypothesis testing is then constructed considering $H_{0i}: \mu_i = 0$ with two-sided alternative, where $\mu_i$ is the mean $\sum_{j =1}^{J} \boldsymbol{\hat{\beta}}_{\text{vocal }j} -  \boldsymbol{\hat{\beta}}_{\text{non-vocal }j}/J$ computed for each voxel $i  = 1, \dots, m$, as described in Equation \eqref{eq:onesam}.

In concordance with results from earlier studies \citep{Pernet1, OLSON2002129, calvert1997activation}, we found activation in the Frontal Pole (FP), Cingulate Gyrus (CG), Superior Frontal Gyrus (SFG), Temporal Occipital Fusiform Cortex (TOF), Lateral Occipital Cortex (LO), Lingual Gyrus (LG), Occipital Fusiform Gyrus (OFG), Inferior Temporal Gyrus (ITG), Supramarginal Gyrus (SG), Angular Gyrus (AG), Superior Temporal Gyrus (STG), Planum Temporale (PT), Middle Temporal Gyrus (MTG), Heschl's Gyrus (HG), Precentral Gyrus (PrG), Thalamus (T), Inferior Frontal Gyrus (IFG), Insular Cortex (I), Central Opercular Cortex (CO), and Frontal Medial Cortex (FM). While our method allows any method for forming clusters, we started from a map computed using RFT with a cluster-forming-threshold equalling $|T_i|>3.2$. This threshold is quite liberal, and therefore we will make additional inferences inside these clusters with a threshold of $|T_i|>4$.

Table \ref{tab:Auditory} includes the lower bounds of the proportion of active voxels ($\bar{\pi}(S) = \bar{a}(S)/|S|$), the size of the cluster ($|S|$), the FWER-corrected $p$-values ($p_{FWER}$) from classical cluster analysis and the mm coordinates of the maximum. The FWER $p$-values based on the clusterwise RFT are reported only for the first cluster-forming-threshold equals to $|T_i|>3.2$, since this method does not allow double-dipping. The results are computed using the permutation-based ARI with the Simes and AORC family using $\delta = 1$. We compared these methods with the original parametric ARI calculated using the R package \texttt{ARIbrain} \citep{ARIbrain}.

\begin{figure}
	\centering
	\includegraphics[width=7cm,height=7cm]{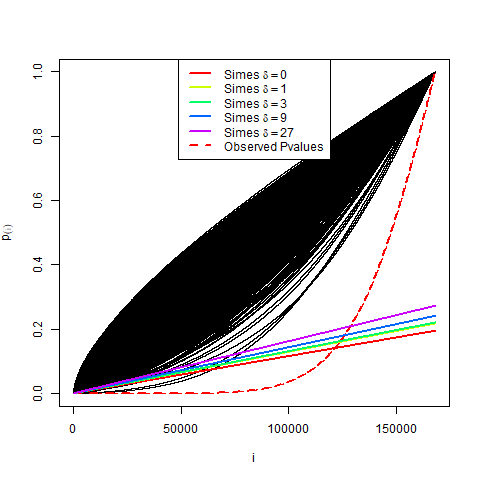}
	\caption{Auditory data: $p$-values null distribution (black lines plus dotted red one) with critical vectors from Simes family considering $\delta \in \{0, 1, 3, 9, 27\}$ (solid colored lines). The red dotted line represents the observed p-values.}
	\label{fig:Auditory}
\end{figure}

As can be seen, the permutation-based ARI, i.e., columns \ref{eq:tblSimesA} for the Simes family and \ref{eq:tblAORCA} for the AORC family in Table \ref{tab:Auditory}, has a better performance overall than the parametric approach, i.e., column \ref{eq:tblparA} in Table \ref{tab:Auditory}. However, the two families of candidate curves return very similar results; this likely reflects the similar structure of these two families of critical vectors. We also applied the shifted versions with $\delta >1$ (see Appendix \ref{auditory_plus} for the results), but found that the loss of power in small clusters is not sufficiently offset by the gain in power in the larger clusters. We believe that this is due to the conservativeness of the null $p$-values as shown in Figure \ref{fig:Auditory}. The family of critical vectors based on the Higher Criticism provide lower TDP than the ones given by the Simes and AORC families, and we put the results in Appendix \ref{auditory_plus}. The family based on the Beta quantile instead does not work on fMRI data due to the large number of variables that make the beta parameters unmanageable in terms of numerical precision. In addition, we believe that the weakness of both these two families is due also to mismatch between the design of the curves based on independent $p$-values that contrasts with a high correlation in the actual data.

\begin{figure}
	\centering
	\includegraphics[width=.8\textwidth]{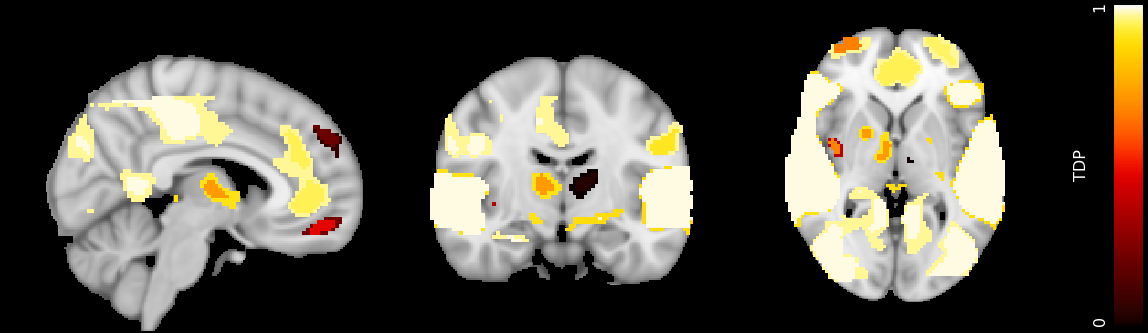}
	\caption{Auditory data: True Discovery Proportion map using the Simes family of critical vectors with $\delta = 1$. Colors express the True Discovery Proportion for clusters based on a threshold of $3.2$ and ``drilled'' down at $4$.}
	\label{fig:Auditory_tdp}
\end{figure}

Figure \ref{fig:Auditory_tdp} shows the TDP bounds as a cluster brain map using the results using the Simes family confidence bound. In these maps, the user can directly interpret activation as the proportion of truly active voxels inside a cluster.
\begin{table}
	\centering
	\caption{Auditory data: Clusters $S$ identified with threshold $\mathbf{|T|}>3.2$ and Active Proportion Percentage $\bar{\pi}(S)$ using Simes and AORC families ($\delta = 1$) and parametric ARI, ``drill down'' clusters at $\mathbf{|T|}>4$. The size of the clusters $|S|$ and the voxel coordinates $(x,y,z)$ are reported for each cluster.}    
	\label{tab:Auditory}
		\begin{adjustbox}{max width=\textwidth}
		\begin{tabular}{@{}rlrrrrcccc@{}}
			\hline
			\multicolumn{1}{c}{Cluster} &\multicolumn{1}{c}{Threshold} & \multicolumn{1}{c}{Size} &\multicolumn{3}{c}{$\%$ active}& \multicolumn{1}{c}{RFT} & \multicolumn{3}{c}{Voxel}     \\
			&  &  && & &\multicolumn{1}{c}{P-Values}& \multicolumn{3}{c}{Coordinates}     \\
			\multicolumn{1}{c}{$S$}&\multicolumn{1}{c}{$t$}  &\multicolumn{1}{c}{$|S|$} & \multicolumn{3}{c}{$\bar{\pi}(S)$}&\multicolumn{1}{c}{$p_{FWER}$}  &   \multicolumn{1}{c}{x} & \multicolumn{1}{c}{y} & z \\ 
			&  &&Perm &Perm & Parametric & &   &  &  \\ 
			&&  &Simes \refstepcounter{equation}(\theequation)\label{eq:tblSimesA}&AORC \refstepcounter{equation}(\theequation)\label{eq:tblAORCA} &Simes \refstepcounter{equation}(\theequation)\label{eq:tblparA}  &  &     &  &\\
			\hline
			\rowcolor{lightgray}
			FP/CG/SFG/TOF/LO &$3.2$ &$40094$  &   $96.77\%$ &  $96.79\%$ & $84.98\%$ &$<0.0001$& -30	&-34&	-16 \\
			\rowcolor{lightgray}
			LG/OFG/ITG/SG/AG & &  &    &   &  &&  &  &   \\
			Left LO/TOF &$4$ &$8983$  &   $99.14\%$ &  $99.14\%$ & $97.66\%$ &$-$ &-30&	-34&	-16 \\
			Right LO/LG/ITG & $4$ &  $7653$  & $98.96\%$&  $98.96\%$ &$97.25\%$ & $-$& 28&	-30&	-18 \\
			Left SFG/FP & $4$ &  $1523$  & $94.75\%$&  $94.81\%$ &$86.28\%$ & $-$&-28&	34&	42 \\
			CG & $4$ &  $1341$  & $94.11\%$&  $94.11\%$ &$84.41\%$ & $-$& 6	&40	&-2 \\
			Right FP & $4$ &  $1327$  & $93.97\%$&  $93.97\%$ &$84.32\%$ & $-$& 30&	56&	28\\
			Left SG/AG & $4$ &  $859$  & $90.69\%$&  $90.8\%$ &$75.79\%$ & $-$& -50&	-56&	36 \\
			Right FP & $4$ &  $243$  & $67.08\%$&  $67.08\%$ &$43.21\%$ & $-$& 30&	64&	-4 \\
			Left SFG & $4$ & 	$202$  & $61.88\%$&  $61.88\%$ &$40.1\%$ & $-$& -18	&8	&52 \\
			Right SFG & $4$ & $122$  & $46.72\%$&  $46.72\%$ &$19.67\%$ & $-$&22&	10	&52  \\ 
			\rowcolor{lightgray}
			Right STG/PT/MTG & $3.2$ &$12540$  &   $90.02\%$& $90.05\%$ & $83.49\%$ &$<0.0001$ &60	&-10&	0 \\
			\rowcolor{lightgray}
			HG/PrG/T & &  &  &  & &  &  &  & \\
			STG/PT/MTG/HG &$4$ &$9533$  &   $99.19\%$ &  $99.19\%$ & $97.8\%$ &$-$& 60&	-10	&0 \\
			PrG & $4$ &  $485$  & $86.19\%$&  $86.19\%$ &$78.35\%$ & $-$& 52&	0&	48 \\
			T & $4$ &  $292$  & $72.6\%$&  $72.6\%$ &$53.77\%$ & $-$& 10&	-10&	8 \\
			\rowcolor{lightgray}
			Left STG/PT/MTG/&  $3.2$ &  $10833$  &     $88.4\%$&  $88.45\%$ & $80.41\%$ &$<0.0001$& -60 &	-12&	2 \\
			\rowcolor{lightgray}
			HG/IFG/T & &  &  &  & &  &  &  & \\
			HG/PT/MTG/STG &$4$ &$7894$  &   $98.99\%$ &  $98.99\%$ & $97.35\%$ &$-$ &-60&	-12&	2 \\
			IFG & $4$ &  $667$  & $88.01\%$&  $88.16\%$ &$74.06\%$ & $-$& -40&	14&	26 \\
			T & $4$ & $34$  & $26.47\%$&  $26.47\%$ &$17.65\%$ & $-$& -14&	-26&	-4 \\
			\rowcolor{lightgray}
			Right IC/CO & $3.2$ &  $408$  & $37.25\%$&  $37.26\%$ &$24.01\%$ &$0.0002$ & 38	&-2&	16 \\
			- & $4$ & $226$  & $67.26\%$&  $67.26\%$ &$43.36\%$ & $-$& 38&	-2&	16 \\ 
			\rowcolor{lightgray}
			Left PrG & $3.2$ &  $276$   &     $49.64\%$&  $49.64\%$  & $43.84\%$  &$0.002$&-52&	-6&	50 \\
			- & $4$ &  $192$  & $71.35\%$&  $71.35\%$ &$63.02\%$ & $-$& -52&	-6&	50 \\ 
			\rowcolor{lightgray}
			FM& $3.2$ &  $270$   &     $22.59\%$&  $22.59\%$  & $13.33\%$  &$0.002$&4&	50&	-14 \\
			- & $4$ & $128$  & $47.66\%$&  $47.66\%$ &$28.13\%$ & $-$& 4&	50&	-14 \\ 
			\rowcolor{lightgray}
			SFG& $3.2$ &  $187$   &     $6.95\%$&  $6.95\%$  & $0\%$  &$0.0123$&6&	52&	38 \\
			- & $4$ & $64$  & $20.31\%$&  $21.23\%$ &$0\%$ & $-$& 6&	52&	38 \\
			\rowcolor{lightgray}
			Left T& $3.2$ &  $176$   &     $1.14\%$&  $1.14\%$  & $0\%$  &$0.0157$&-14	&-14&	10 \\
			- & $4$ & $49$  & $4.08\%$&  $4.08\%$ &$0\%$ & $-$& -14&	-14	&10 \\
			\hline
		\end{tabular}
			\end{adjustbox}
\end{table}

\subsection{Rhyme Data}\label{food-data}

Subsequently, we analyzed data from $13$ subjects making rhyming judgments for pairs of either words or pseudo-words, collected by \cite{Rhyme} and available at \url{https://openneuro.org/datasets/ds000003/versions/1.0.0}. The analysis follows directly the one performed in Section \ref{auditory-data}, but the neural activation during the word stimulus was analyzed. 

We found activity in Paracingulate Gyrus (PG), Lateral Occipital Cortex (LOC), Superior Frontal Gyrus (STG), Frontal Operculum Cortex (FOC), Putamen (P), Inferior Frontal Gyrus (IFG), Lingual Gyrus (LG), Occipital Fusiform Gyrus (OFG), Insular Cortex (IC), Cingulate Gyrus (CG), Superior Pariental Lobe (SPL), and Post Central Gyrus (PCG) \citep{Rhyme1}. The cluster map is thresholded the same as the previous dataset: using cluster-wise RFT with a threshold of $|T_i|>3.2$. We then drilled down  $\bar{\pi}(S)$ using a threshold of $|T_i|>4$. \angela{For completeness, we also analyzed the clusters defined by the threshold-free cluster enhancement (TFCE) method \citep{smith2009threshold}. The analysis results are reported in Appendix \ref{rhyme_plus}.}

Figure \ref{fig:Food} shows the null distribution of the $p$-values from the one-sample t-test (two-sided alternative) for the contrast regarding the word stimulus. As in Section \ref{auditory-data}, the Table \ref{tab:Food} represents the results using the Simes family in column \ref{eq:tblSimes11}, and the AORC family in column \ref{eq:tblAORC11} with $\delta$ equalling $27$. As can be seen, the power improvement over the parametric method, i.e., column \ref{eq:tblpar11} of Table \ref{tab:Food}, is striking in this dataset. We decided to take $\delta$ equals $27$ since, in this case, all clusters based on RFT have large sizes. The results using $\delta \in \{0,1,9\}$ are shown in Appendix \ref{rhyme_plus}. Once again, the critical vectors based on Higher Criticism and the beta distribution do not work well due to the high correlation in the data.

\begin{figure}
	\centering
	\includegraphics[width=.8\textwidth]{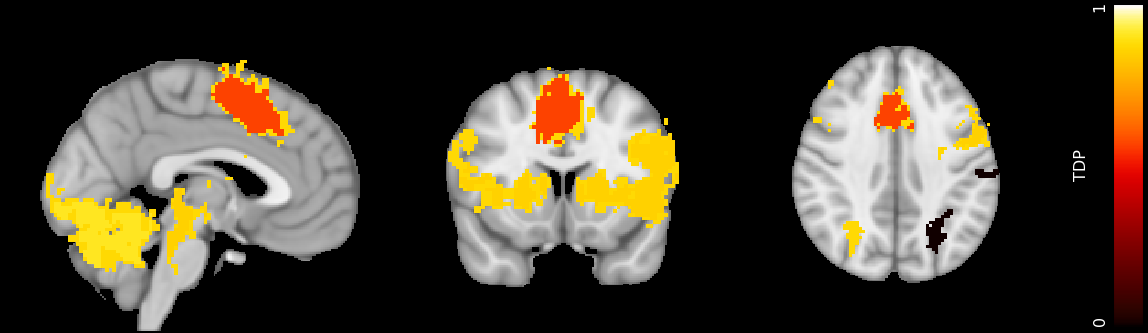}
	\caption{Rhyme data: True Discovery Proportion map using the Simes family of critical vectors with $\delta = 27$. Colors express the True Discovery Proportion for clusters \angela{corresponding} to a threshold of $3.2$ and ``drilled'' down at $4$.}
	\label{fig:Food_tdp}
\end{figure}

Finally, Figure \ref{fig:Food_tdp} shows the TDP represented in Table \ref{tab:Food} as cluster brain map. 
\begin{figure}
	\centering
	\includegraphics[width=7cm,height=7cm]{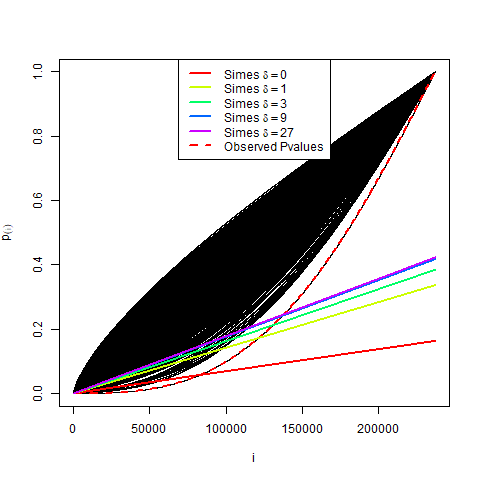}
	\caption{Rhyme data: $p$-values null distribution (black lines plus dotted red one) with critical vectors from Simes family considering $\delta \in \{0, 1, 3, 9, 27\}$ (solid colored lines). The red dotted line represents the observed p-values.}
	\label{fig:Food}
\end{figure}
\begin{table}
	\centering
	\caption{Rhyme data: Clusters $S$ identified with threshold $\mathbf{|T|}>3.2$ and Active Proportion Percentage $\bar{\pi}(S)$ using Simes and AORC families ($\delta = 27$) and parametric ARI, ``drill down'' clusters at $\mathbf{|T|}>4$. The size of the clusters $|S|$ and the voxel coordinates $(x,y,z)$ are reported for each cluster.}      
	\label{tab:Food}
		\begin{adjustbox}{max width=\textwidth}
		\begin{tabular}{@{}rlrrrrcccc@{}}
			\hline
			\multicolumn{1}{c}{Cluster} &\multicolumn{1}{c}{Threshold} & \multicolumn{1}{c}{Size} &\multicolumn{3}{c}{$\%$ active}& \multicolumn{1}{c}{RFT} & \multicolumn{3}{c}{Voxel}     \\
			&  &  && & &\multicolumn{1}{c}{P-Values}& \multicolumn{3}{c}{Coordinates}     \\
			\multicolumn{1}{c}{$S$}&\multicolumn{1}{c}{$t$}  &\multicolumn{1}{c}{$|S|$} & \multicolumn{3}{c}{$\bar{\pi}(S)$}&\multicolumn{1}{c}{$p_{FWER}$}  &   \multicolumn{1}{c}{x} & \multicolumn{1}{c}{y} & z \\ 
			&  &&Perm &Perm & Parametric & &   &  &  \\ 
			&&  &Simes \refstepcounter{equation}(\theequation)\label{eq:tblSimes11}&AORC \refstepcounter{equation}(\theequation)\label{eq:tblAORC11} &Simes \refstepcounter{equation}(\theequation)\label{eq:tblpar11}  &  &     &  &\\
			\hline
			\rowcolor{lightgray}
			LOC/LG/OFG/PG/SFG & $3.2$ & $34115$  &  $89.15\%$ & $89.4\%$ &  $38.16\%$ & $<0.001$ & $4$ &  $12$ &  $48$ \\
			\rowcolor{lightgray}
			FOC/P/IFG/IC/CG & &  &    &   &  & &  &  &   \\
			LOC/LG/OFG & $4$ &$11045$ &  $91.21\%$ & $91.45\%$ &  $42.01\%$& $-$ & $-6$ &  $-56$ &  $-12$ \\
			FOC/P/IFG/IC &$4$ & $6930$   & $85.75\%$ & $86.2\%$ &  $29.32\%$& $-$ & $-42$ &  $14$ &  $-6$ \\
			PG/SFG/CG & $4$ &  $2100$  & $57\%$ & $57.81\%$ & $18.05\%$& $-$ & $4$ & $12$ &  $48$ \\
			Left P & $4$ &  $38$  & $2.63\%$ & $2.63\%$ & $2.63\%$& $-$ & $-32$ &  $-18$ &  $-8$ \\
			\rowcolor{lightgray}
			Left SPL/PCG & $3.2$ & $1546$ & $1.49\%$ & $1.75\%$ &  $0\%$ & $<0.001$ & $-24$ &  $-62$ &  $44$ \\
			\hline
		\end{tabular}
			\end{adjustbox}
\end{table}

\section{Validating Permutation-based ARI}\label{validiting-and-power-of-the-permutation-based-ari-inference}

\angela{FMRI data has noise characteristics that are hard to simulate using parametric distributions. Therefore, when performing simulations often resting-state fMRI data (i.e. fMRI data with no stimulus linked BOLD signal) is used. In these} null data\angela{,} the hypothesis of mean zero activation between groups is true,  \angela{while still retaining the noise characteristics of fMRI data}. \cite{Eklund} found that many software programs such as FSL \citep{Jenkinson} and SPM \citep{SPM} do not \angela{properly} control the probability or the average proportion of the false positives in cluster-wise inference \angela{when RFT assumptions are not met}. \angela{As for ARI RFT assumptions do not have to be met}, we want to analyze the false positive rate of the permutation-based ARI using resting-state fMRI data with no signal. \angela{For this we used the }Oulu data\angela{set} provided by the $1000$ Functional Connectomes Projects \citep{Biswal}. The pre-processing pipeline follows the one used in \cite{Eklund}. In particular, we analyzed the Oulu data\angela{set} from \url{https://tinyurl.com/clusterfailure} considering fMRI images pre-processed by FSL \citep{Jenkinson} with a level of smoothness equal to $6$mm FWHM and $6$ different first level designs (four event activity paradigms, and two block activity paradigms).

The Oulu data\angela{set} consists of $103$ subjects; however, to estimate the false positive rate, this set of subjects is not sufficient. In addition, \cite{Eklund} found asymmetric errors in the case of permutation test for the one-sample t-test using the Oulu data; therefore, we validate the permutation-based and parametric ARI, performing the two-sample t-tests. We select two groups of $20$ subjects by randomly permuting $100$ times the subject numbers and selecting the first $40$ of this permuted dataset. \cite{Eklund} underlines that the estimate of the familywise false positive rate is unbiased, even if these random datasets are not independent. Finally, the set of voxels used as a cluster map is \angela{used as }the whole-brain mask. Please see Appendix \ref{eklund_plus} for the results the one-sample t-tests.

\begin{figure}
	\centering
	\includegraphics[width=.8\textwidth]{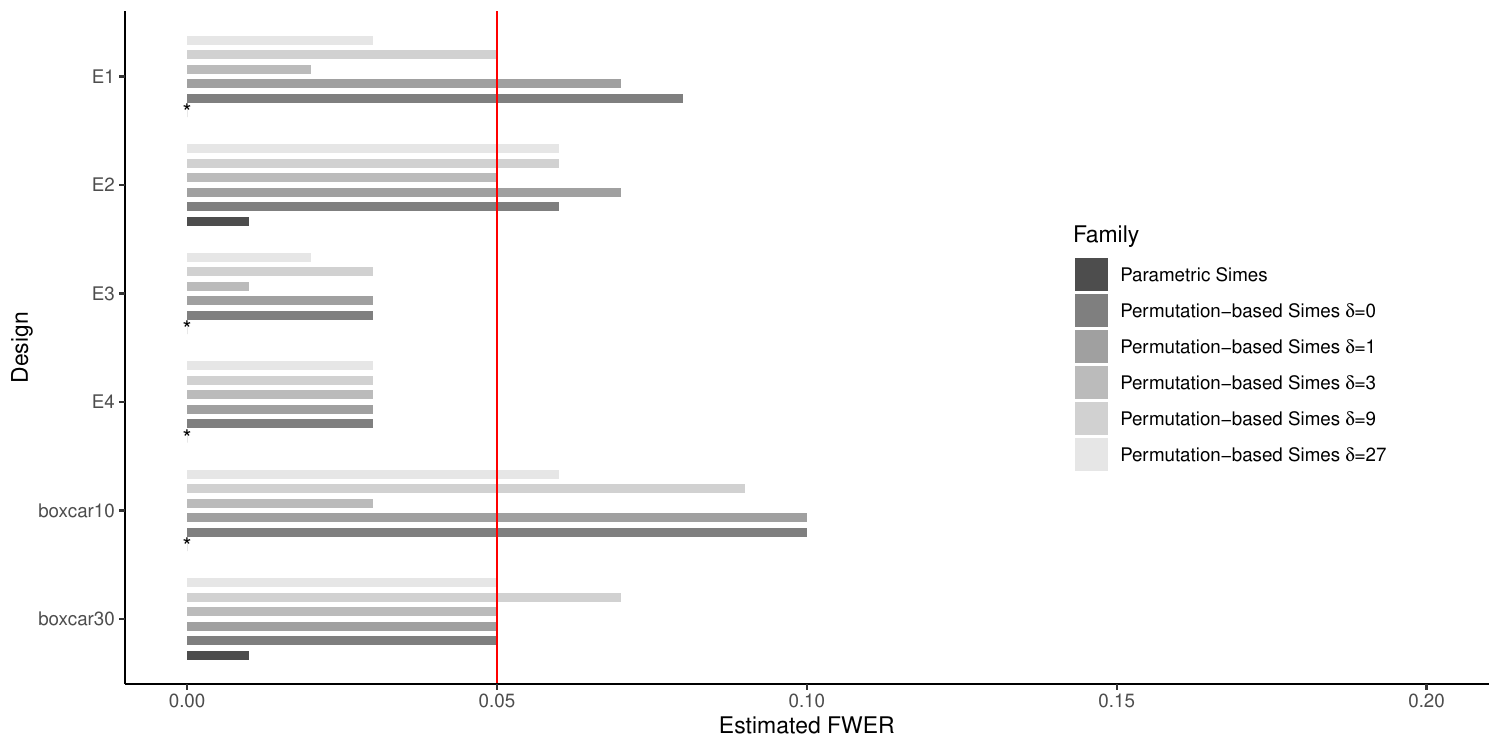}
	\caption{Estimated FWER considering six different first level designs, i.e., two-block activity paradigms: boxcar10 (10-s on-off), boxcar30 (30-s on-off), and four event activity paradigms, i.e., E1 (single event of 2-s activation, 6-s rest), E2 (single event 1- to 4-s activation, 3- to 6-s rest, randomized), E3 (13 events of 3–6 s for each task), and E4 (13 events of 3–6 s for each task, randomized), and six different methods to compute the TDP's lower bound (parametric Simes and permutation-based Simes considering five different values of the shift parameter, i.e., $\delta \in \{0, 1, 3, 9, 27\}$). The solid red lines represents \angela{the estimated }nominal FWER equals $0.05$, \angelaa{while the star symbols describe estimated FWER equals $0$.}}
	\label{Oulu}
\end{figure}

Figure \ref{Oulu} shows the FWER estimated considering six different first level designs (i.e., two-block activity paradigms: boxcar10 (10-s on-off), boxcar30 (30-s on-off) and four event activity paradigms: E1 (single event of 2-s activation, 6-s rest), E2 (single event 1- to 4-s activation, 3- to 6-s rest, randomized), E3 (13 events of 3–6 s for each task), and E4 (13 events of 3–6 s for each task, randomized). See \cite{Eklund} and \cite{eklund2019cluster} for more details about tested parameter combinations.

To sum up, in most cases, the parametric-based ARI returns a false positive rate equal to $0$, while the permutation-based ARI with the Simes family returns false positive rates greater than $0$. Considering the boxcar10, E1, and E2 designs, the families with lower shifts, i.e., $0$ and $1$, are more powerful than imposing the shift equals $3$, $9$, and $27$. Therefore, both methods (i.e., parametric-based and permutation-based ARI) control the FWER. \angela{In addition, the analysis confirms the conservativeness of the parametric ARI method in case of strong positive dependence due to the Simes inequality and positive regression dependence on subsets (PRDS) \citep{Sarkar, Neuvial3} assumptions.}

\section{Simulation study}\label{Simulation}
We simulate data considering the simple following model \angela{(i.e., model \eqref{one-sample-model})}:
\begin{align*}
	\angela{D_{ij} = \mu_i + \epsilon_{ij}^\star}
\end{align*}
where \angela{$\mathbf{D}_j \in \mathbb{R}^m$}, with $j = 1, \dots, J$, $J$ is the number of independent observations \angela{(i.e., subjects)} and $m$ is the total number of \angela{voxels}. The noise \angela{$\epsilon_j^\star \in \mathbb{R}^m$} follows the multivariate normal distribution with mean $0$ and spatial correlation structure, i.e., \angela{$\epsilon_j^\star \sim \mathcal{N}(0,\Sigma_{\theta})$, where $\theta$ describes how rapidly the correlation declines with respect to the distance between two voxels. The three-dimensional coordinates of the voxels are defined as all combinations of vector $c = \{1, \dots, m^{1/3}\}$, then $\Sigma_{\theta} = \exp(- \theta K)$ where $K$ is the matrix containing the euclidean distances between the three-dimensional coordinates' voxels. For example, if $\theta = 0.2$, the correlation between two voxels with a distance of $1$ equals $0.819$, while the the correlation between two voxels with a distance of $5$ equals $0.368$, and so on.} The signal \angela{$\boldsymbol{\mu} \in \mathbb{R}^{m}$} is computed considering the difference in means having power of the one-sample t-test equals $0.8$, \angela{i.e., $\boldsymbol{\mu} = (z_{1-\alpha/2} + z_{1-\beta})/\sqrt{J}$, where $\alpha  = 0.05$ is the significance level, $\beta = 0.8$ is the power level, and $z_{a}$ is the quantiles of the standard normal distribution at level $a$}. The signal $\boldsymbol{\mu}$ is equal to $0$ under the null hypothesis. 

\begin{figure}
	\centering
	\includegraphics[width=.7\textwidth]{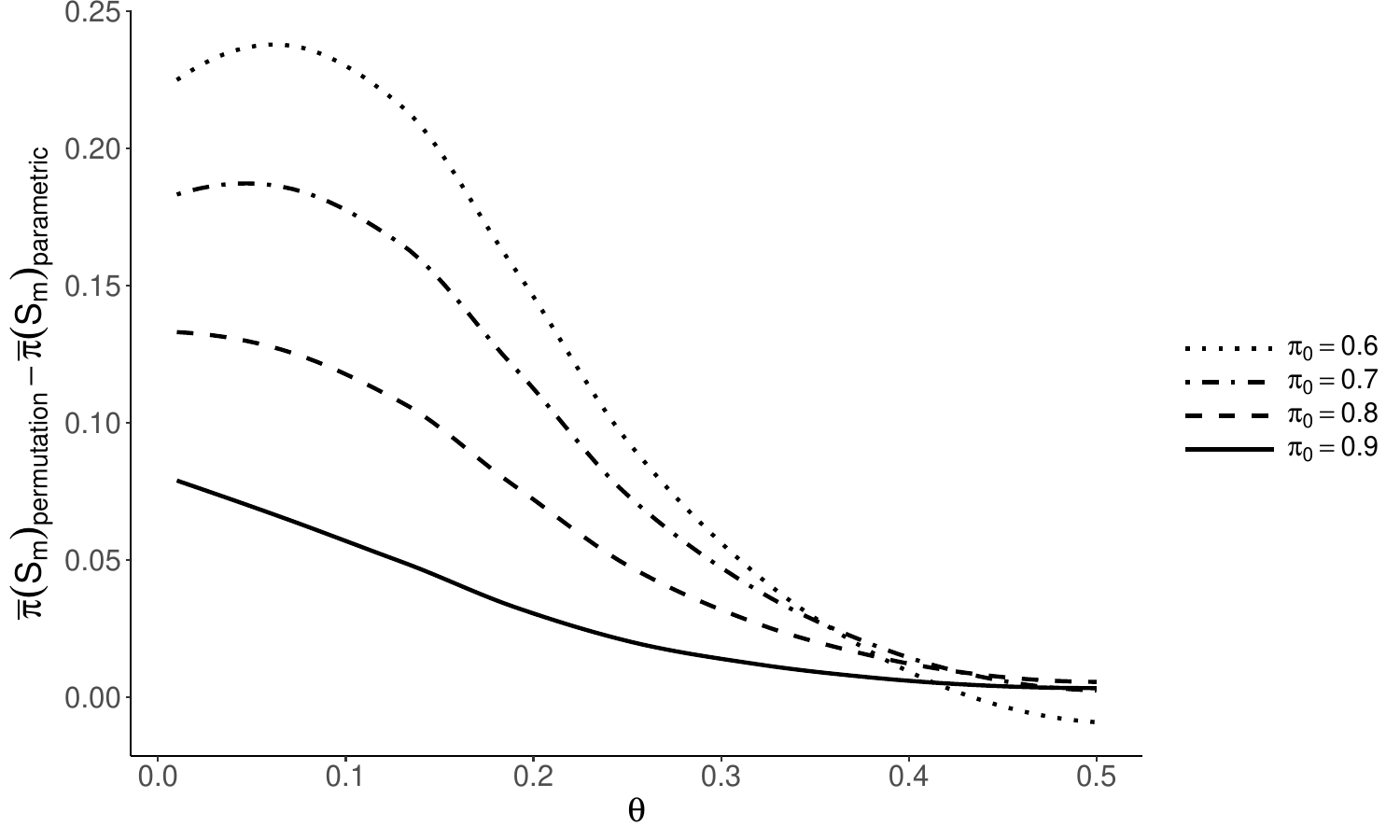}
	\caption{Difference of lower bounds for the true discoveries proportion considering the permutation $\bar{\pi}(S_m)_{\text{permutation}}$ and parametric $\bar{\pi}(S_m)_{\text{parametric}}$ methods using simulated data and considering the full set of hypotheses $S_m$ over different values of \angela{$\theta \in \{0,0.01, \dots, 0.5\}$} and $\pi_0 \in \{0.6, 0.7, 0.8, 0.9 \}$.}
	\label{sim:pi}
\end{figure}

First of all, we want to understand how the improvement of the nonparametric TDP lower bound changes concerning \angela{$\theta$} and the proportion of null hypotheses $\pi_0$. Let $J = 50$, $m = 1000$, \angela{$\theta \in \{0,0.01, \dots, 0.5\}$} and $\pi_0 \in \{0.6, 0.7, 0.8, 0.9\}$, we simulate data $1000$ times and the mean of $\bar{\pi}(S_m)$ over simulation is represented. The Simes family of confidence bound without shift is taken into account to compare with the parametric approach directly. Having no prior knowledge about the structure of the set of hypotheses to analyze, we consider the full set of hypotheses, i.e., $S_m$. Figure \ref{sim:pi} shows the difference of $\bar{\pi}(S_m)$ computed using the permutation and parametric methods over the \angela{$\theta$} and $\pi_0$ values. As expected, the permutation approach gets some power with respect to the parametric one in the case of correlation between pairs of variables. It can handle any type of dependence structure of the $p$-values.

\begin{figure}
	\centering
	\includegraphics[width=.7\textwidth]{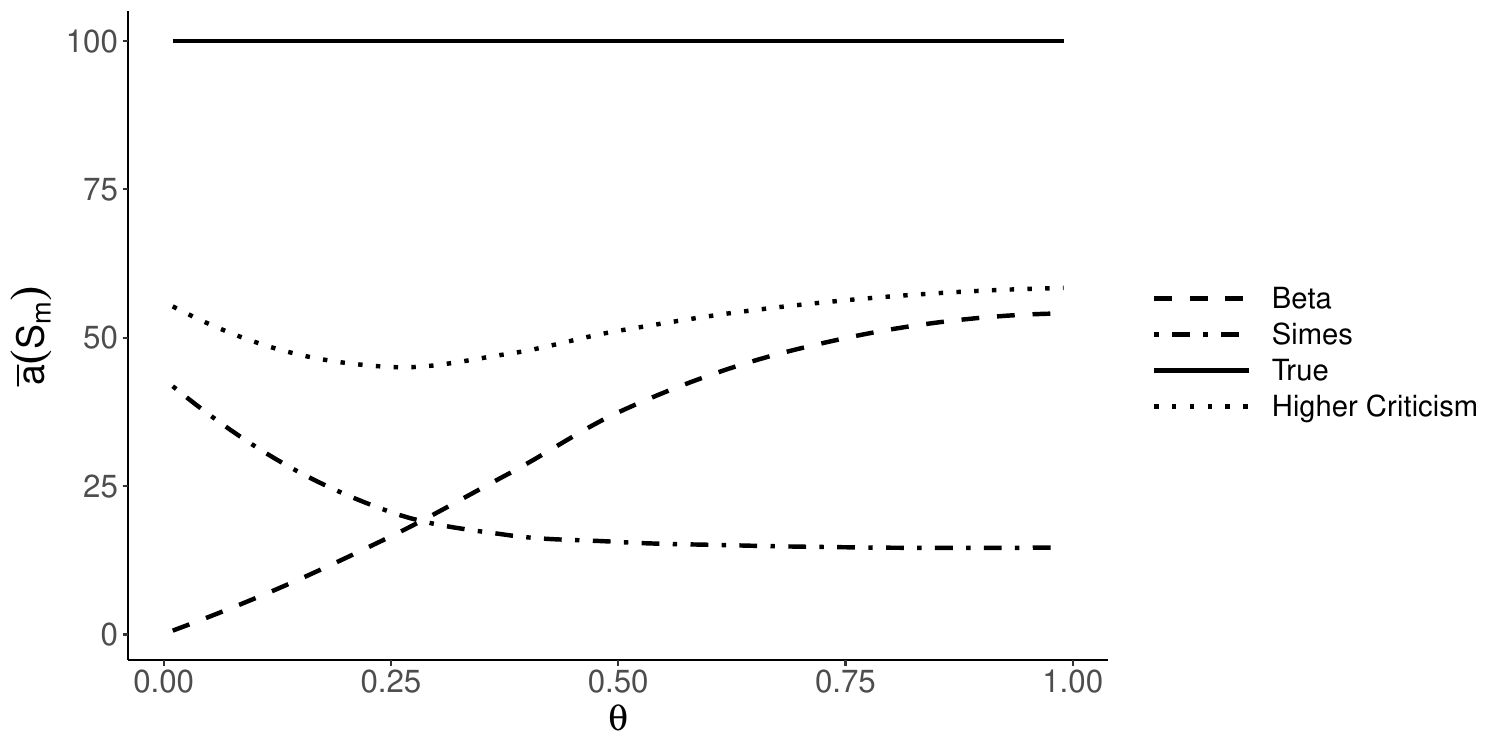}
	\caption{Simulated True Discovery lower bound over $S_m$ and different values of \angela{$\theta \in \{0,0.01, \dots, 0.99,1\}$} using the Higher Criticism (dotted line), Beta (dashed line) and Simes critical vectors (dotted dashed line). The solid line represents the number of true discoveries which equals $100$ considering $1000$ variables and the proportion of null hypotheses $\pi_0 = 0.9$.}
	\label{sim:hc_beta}
\end{figure}

Secondly, we want to examine why certain families of critical curves do not provide good results in Section \ref{applications}. The Higher Criticism critical vector \eqref{hc}, the Beta critical vector \eqref{beta} and the Simes critical vector \eqref{eq1} are then used to computed $\bar{a}(S_m)$ using simulated data with $\pi_0= 0.9$, $m = 1000$ and $J = 50$. As previously, we repeat the simulations $1000$ times for each framework, and the mean value of $\bar{a}(S_m)$ is computed. Figure \ref{sim:hc_beta} shows the behaviour of these three families of critical vectors respect to \angela{$\theta \in \{0,0.01,\dots, 0.99,1\}$}. In Section \ref{lambdaCal}, we said that the Higher Criticism and Beta families could be problematic in the case of a strong correlation between tests. As expected, the Beta critical vector does not work in the case of a strong correlation between variables, \angela{i.e., low value of $\theta$}. \angela{However, the Higher Criticism family seems to work considering various values of $\theta$ in contrast to the results with fMRI data. This may be due to the different spatial correlation in the fMRI data, which is much more complex than that specified in the simulations. However, it can be seen that the lower bound for the TDP calculated by the Higher Criticism family is close to the one computed by the Simes family as the correlation increases. Finally, the Beta family works only if $\theta>0.2$ (i.e., correlation between voxels equals $0.28$ on average). However, high values of correlation are unrealistic in real applications. For example, the mean correlation across $10,000$ randomly sampled voxels equals $0.25$ in the case of Rhyme data.}

\begin{figure}[!htb]
	\centering
	\includegraphics[width=.9\textwidth]{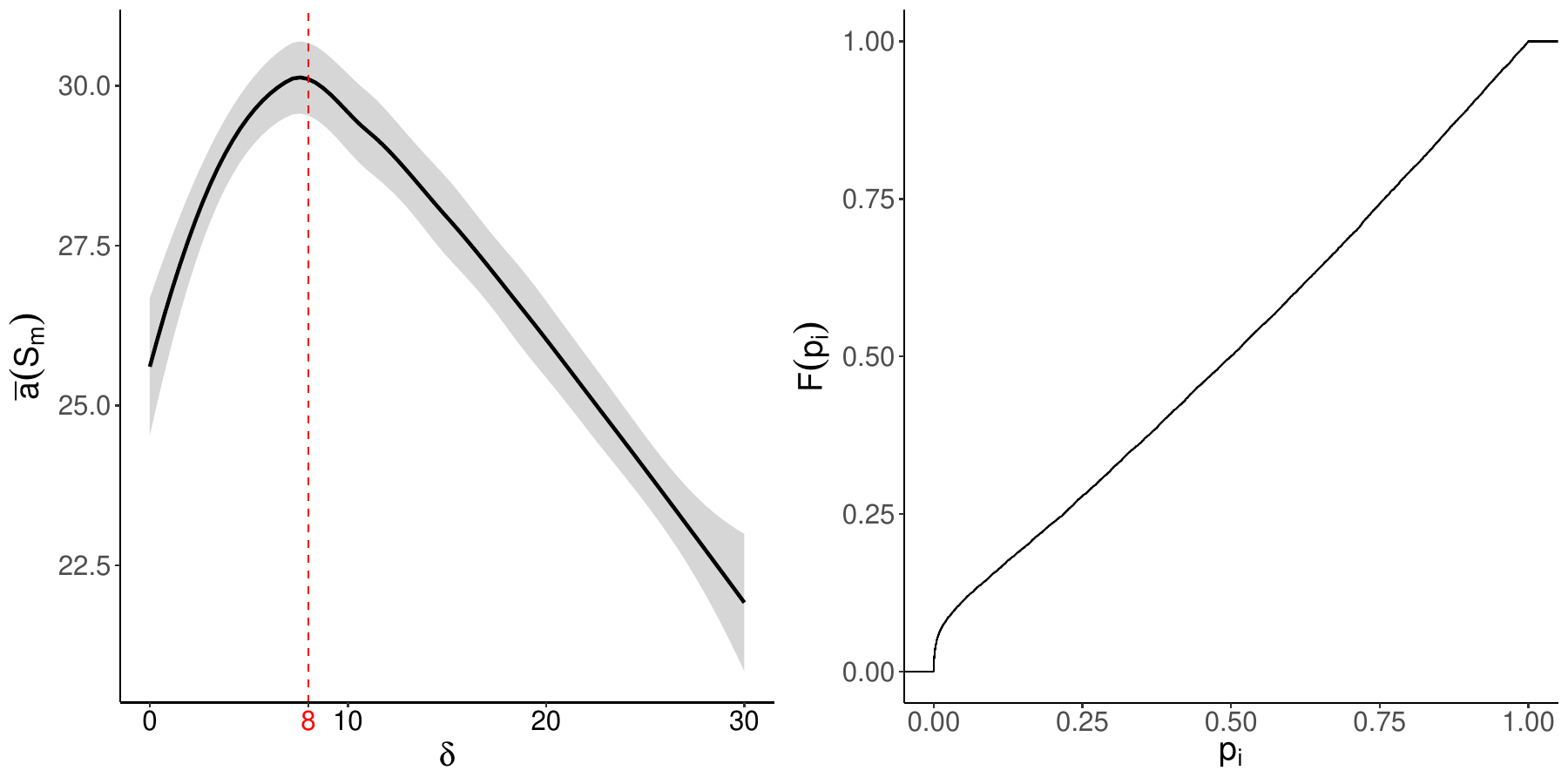}
	
	\caption{ Left side: True Discovery Lower Bound using simulated data. The full set of hypotheses $S_m$ is considered over different values of $\delta$. Right side: Empirical Cumulative Density Function of observed raw $p$-values, i.e, $F(p_i)$.}
	\label{sim:pv}
\end{figure}

Thirdly, we want to analyze how the Simes family of critical curves \eqref{eq1} works if anti-conservative $p$-values distribution is considered. Let $J=50$, $m=1000$, \angela{$\theta = 0.2$ (i.e., correlation between voxels equals $0.28$ on average)}, and $\pi_0=0.9$, we compute $\bar{a}(S_m)$ for each $1000$ simulations, and once again the mean over simulations is reported. Figure \ref{sim:pv} shows $\bar{a}(S_m)$ considering the Simes family using $\delta \in \{0, \dots, 30\}$. We can note that the shifted version works well in the case of anti-conservative $p$-values if the corrected value for the tuning parameter $\delta$ is chosen, described by the red dotted line, i.e., $\delta = 8$. 

\begin{figure}
	\centering
	\includegraphics[width=.9\textwidth]{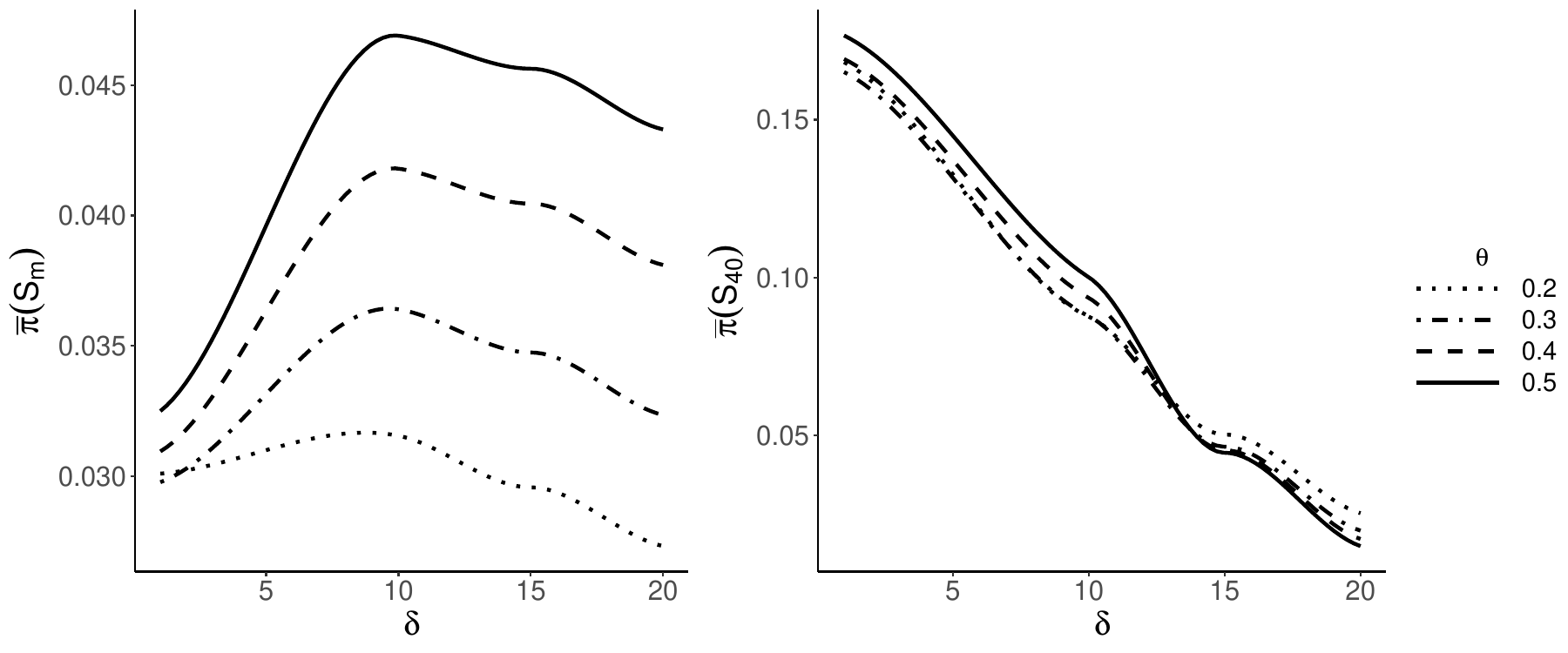}
	\caption{Lower Bounds for the True Discovery proportion using simulated data with \angela{$\theta \in \{0.2, 0.3, 0.4, 0.5\}$}. In the left figure, the full set of hypotheses is considered, while in the right figure a random sample of $40$ hypotheses is analyzed. The critical vectors based on the Simes family with $\delta \in \{0, 5, 10, 15, 20\}$ are used in both situations.}
	\label{sim:deltas}
\end{figure}

Therefore, we explore how the Simes family of critical curves \eqref{eq1} works with different values of $\theta$ and $S$ size. The left part of Figure \ref{sim:deltas} shows the mean of the lower bounds for the true discoveries considering the full set of hypotheses, i.e., $S_m$,  and $\delta \in \{0,5,10,15,20\}$ over $1000$ simulations. We can note that in almost all scenarios, the shifted version outperforms the unshifted ones. The difference gets smaller if \angela{$\theta$ decreases (i.e., the correlations between voxels increase)}. However, the situation changes if we compute the TDP for a smaller set of hypotheses than $S_m$ as shown in the right part of Figure \ref{sim:deltas}. In this case, we randomly sample $40$ hypotheses from the false null ones, i.e., $S_{40}$.

\begin{figure}
	\centering
	\includegraphics[width=.9\textwidth]{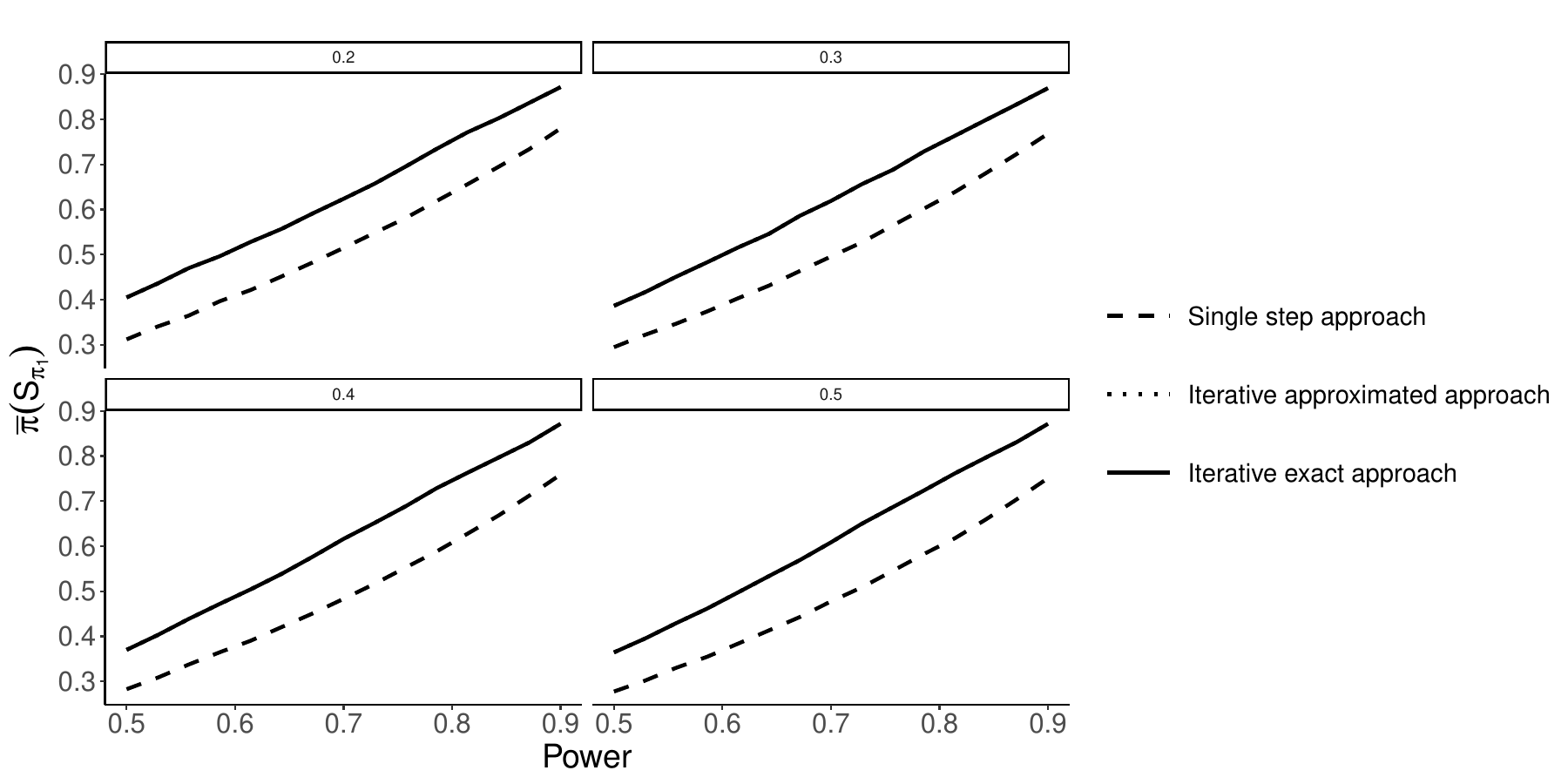}
	\caption{Simulated True Discovery proportion lower bound for $S_{\pi_1}$ over different values of \angela{$\theta$} and power using the single-step (dashed line), iterative approximated version (dotted line) and iterative exact version (solid line). \angela{The dotted line is behind the solid line.}}
	\label{iterative_sim}
\end{figure}

Finally, the performance of the iterative approach proposed in Section \ref{iterative} is compared with respect to the single-step one presented in Section \ref{methods}. Let consider directly $S_{\pi_1}$ the set of true discoveries, therefore $\pi(S_{\pi_1}) = 1$. Figure \ref{iterative_sim} shows the true discovery proportion $\bar{\pi}(S_{\pi_1})$ computed on $1000$ simulated datas with $\pi_0$ equals $0.9$, \angela{$\theta \in \{0.2, 0.3, 0.4, 0.5\}$} and different levels of power used to simulate the data. In this case, we consider $m = 64$ so that we can use the exact iterative method. First of all, we can see how the approximated iterative version equals the exact one, and, more important, how both of them uniformly improve the single-step approach.

\angela{To sum up, we suggest using the Higher Criticism and Beta families if the correlation across the variables is supposed to be low. Besides, we recommend considering the shifted version of the Simes or AORC families if the interest is in large sets of hypotheses rather than small ones. The shifted version of the Simes and AORC families is again recommended if the p-values' distribution is expected to be anti-conservative. We stress that the value of the $\delta$ parameter must be decided a priori and chosen reasonably concerning the data analyzed, as seen in the fMRI data application. }

We include some simulation analyses to examine the power of the iterative approach presented in Section \ref{iterative} and the influence of the number of combinations chosen for its approximated version in Appendix \ref{it_plus}. \angela{Finally, we show some simulation study in Appendix \ref{equi-corr} in the case of equi-correlation variance structure for $\epsilon_j^\star$.}

\section{Conclusions}
Our proposed method finds simultaneous lower bounds for the TDP over all possible hypothesis subsets using the permutation theory in a computationally efficient way. As a simultaneous method, it allows the decision of which hypothesis sets to analyze to be entirely flexible and post-hoc, that is, the user can choose it after seeing the data and revise the choice as often as he/she wants. It is particularly useful in fMRI single and multi-subjects analysis to infer inside clusters, resolving the so-called spatial specificity paradox, without falling into the double-dipping problem.

\angelaa{A method that has some apparent similarity with the permutation-based ARI approach is the TFCE approach proposed by \cite{smith2009threshold}. Both methods use permutation approaches, and both are flexible in the choice of threshold. There are two important differences between the methods, however.
	First, while TFCE allows data-driven thresholds, our proposed method is more flexible since it allows simultaneous use of many thresholds, which can be chosen after viewing the data. Second---and more importantly---permutation-based ARI provides additional information about the clusters: a lower bound for the proportion of true discoveries. In contrast, TFCE remains a cluster-level inferential method, returning only a $p$-value for the clusterwise null hypothesis.}

Permutation-based methods are recommended whenever they can be used, gaining power over the parametric approaches, especially when the $p$-values are strongly dependent, as for fMRI data. In this work, we used permutation theory to calculate the critical vector needed for ARI. 
Our method adapts to the correlation structure of the data in an exact way, by means of the calibration of the parameter  $\lambda_{\alpha}$.
In this way, our method remedies the existing issues of anti-conservativeness and conservativeness. Indeed, we found that the permutation-based method has more power than the parametric approach both in simulated and real data and confirmed FWER control using resting-state null data.

Permutation methods are not assumption-free but require exchangeability of the test statistics under the null hypothesis. \angelaa{We showed the results using} the OLS one-sample t-test for fMRI group analysis, having a fast and straightforward computation of the permutation null distribution, randomly flipping the sign of each subject's contrast. The exchangeability assumption needed to perform permutation-based methods is satisfied, i.e., the error terms of the model need to be symmetric around $0$. \angelaa{However, the method is also applicable using other statistical tests, e.g., two sample t-tests.} Even if permutations are employed to perform the method proposed, the computation time remains low, e.g., around \angela{$210$} seconds using the single-step method, while around $1$ hour using the iterative approach with $10$ combinations, having $150,000$ hypotheses and $1000$ permutations. The computation time is related to a device with a processor having $1.8$ GHz CPU and $16$ GB of RAM, finally, the R package \texttt{pARI} available on CRAN \citep{ARIperm} based on the C++ language was used.

The proposed method is general, allowing different families of confidence bounds. The choice of the family is critical since it directly influences the bounds for the true discovery proportions, and thus the power properties of the method. \angela{Simulations and real data analysis suggest the Simes \eqref{eq1} and AORC \eqref{eq2} families in the fMRI framework, while the Higher Criticism \eqref{eq3} and Beta families \eqref{beta} if the correlation between variables is supposed to be low. Simes \eqref{eq1} and AORC \eqref{eq2} families depend on the shift parameter $\delta$. We recommend fixing $\delta=1$ if the practitioner is interested in computing the lower bound for the TDP in small clusters, while $\delta>1$ if the attention is focused on large clusters.} Finally, drilling down may increase or decrease the lower bound for the TDP of some subclusters if a large cluster is analyzed, e.g., the first cluster found in Section \ref{food-data}.
This suggestion is also confirmed by the simulation analysis. We found that the shifted versions gain power if the raw $p$-values are anti-conservative. 
Other types of families that we analyzed, based on Higher Criticism \citep{Donoho} and Beta quantiles, do not seem to perform well in fMRI data analysis due to the strong correlation among the voxels, as also illustrated in the simulation study of Section \ref{Simulation}. Generally, we suggest a family of critical vectors more concentrated on small $p$-values if the number of rejected hypotheses may be low, and a family of critical vectors more diffuse if the number of rejected hypotheses may be high. 

Finally, we implement the iterative approach proposed by \cite{Jesse} in the simultaneous post-hoc inference scenario, which uniformly improves the \cite{Neuvial3}'s bounds in most cases. There is a power gain here, but for fMRI data with sparse signal, the gain is small and comes at a large computational cost. 

Our presently used method provides a useful and practical selective inference for fMRI data that exploits the advantages of permutation theory and the closed-testing procedure, resolving the spatial specificity paradox with quite fast computation time. The proposed method would be applicable not only for the fMRI data but more for any other data types that may yield multiple testing problems and cluster-wise inference (e.g., electroencephalography data and genomic data).



\subsection*{Author contributions}

All authors have directly participated in the planning and execution of the presented work.



\subsection*{Conflict of interest}

The authors declare no potential conflict of interests.

\subsection*{Acknowledgments}

Angela Andreella gratefully acknowledges funding from the grant BIRD2020/SCAR ASSEGNIBIRD2020\_01 of the University of Padova, Italy, and PON 2014-2020/DM 1062 of the Ca’ Foscari University of Venice, Italy. Some of the computational analyses done in this manuscript were carried out using the University of Padova Strategic Research Infrastructure Grant 2017: "CAPRI: Calcolo ad Alte Prestazioni per la Ricerca e l’Innovazione", \url{http://capri.dei.unipd.it}. 

\subsection*{Data availability statement}
The fMRI data used are available from \url{http://github.com/angeella/fMRIdata}, and the approach is developed as \texttt{R} package available on CRAN (\url{https://CRAN.R-project.org/package=pARI}).

%
%
%

\appendix

\section{Auditory data analysis}\label{auditory_plus}
Table \ref{tab:Auditory1} contains the results using the single-step approach presented in Section \ref{methods} with $\delta=1$ as Table \ref{tab:Auditory} of Subsection \ref{auditory-data}. Instead Table \ref{tab:AuditorySimesDelta} shows the results using the Simes family of confidence bounds, considering $\delta$ equal respectively to $0$, $9$ and $27$. In the same way, Table \ref{tab:AuditoryFinnerDelta} proposes the results using the AORC family. Finally, Table \ref{tab:AuditoryHCBeta} shows the results using the family of critical vectors based on the higher criticism. In all analysis, we consider $\alpha$ equals $0.05$.

\begin{table}
	\centering
	\caption{Auditory data: Clusters $S$ identified with threshold $\mathbf{|T|}>3.2$ and Active Proportion Percentage $\bar{\pi}(S)$ using Simes and AORC families ($\delta = 1$) and parametric ARI, ``drill down'' clusters at $\mathbf{|T|}>4$. The single-step method is applied. The size of the clusters $|S|$ and the voxel coordinates $(x,y,z)$ are reported for each cluster.}  
	\label{tab:Auditory1}
		\begin{adjustbox}{max width=\textwidth}
		\begin{tabular}{@{}rlrrrrcccc@{}}
			\hline
			\multicolumn{1}{c}{Cluster} &\multicolumn{1}{c}{Threshold} & \multicolumn{1}{c}{Size} &\multicolumn{3}{c}{$\%$ active}& \multicolumn{1}{c}{RFT} & \multicolumn{3}{c}{Voxel}     \\
			&  &  && & &\multicolumn{1}{c}{P-Values}& \multicolumn{3}{c}{Coordinates}     \\
			\multicolumn{1}{c}{$S$}&\multicolumn{1}{c}{$t$}  &\multicolumn{1}{c}{$|S|$} & \multicolumn{3}{c}{$\bar{\pi}(S)$}&\multicolumn{1}{c}{$p_{FWER}$}  &   \multicolumn{1}{c}{x} & \multicolumn{1}{c}{y} & z \\ 
			&  &&Perm &Perm & Parametric & &   &  &  \\ 
			&&  &Simes \refstepcounter{equation}(\theequation)\label{eq:tblSimes}&AORC \refstepcounter{equation}(\theequation)\label{eq:tblAORC} &Simes \refstepcounter{equation}(\theequation)\label{eq:tblpar}  &  &     &  &\\
			\hline
			\rowcolor{lightgray}
			FP/CG/SFG/TOF/LO &$3.2$ &$40094$  &   $96.7\%$ &  $96.73\%$ & $84.98\%$ &$<0.0001$& -30	&-34&	-16 \\
			\rowcolor{lightgray}
			LG/OFG/ITG/SG/AG & &  &    &   &  &&  &  &   \\
			Left LO/TOF &$4$ &$8983$  &   $99.11\%$ &  $99.11\%$ & $97.66\%$ &$-$ &-30&	-34&	-16 \\
			Right LO/LG/ITG & $4$ &  $7653$  & $98.96\%$&  $98.96\%$ &$97.25\%$ & $-$& 28&	-30&	-18 \\
			Left SFG/FP & $4$ &  $1523$  & $94.75\%$&  $94.75\%$ &$86.28\%$ & $-$&-28&	34&	42 \\
			CG & $4$ &  $1341$  & $94.11\%$&  $94.11\%$ &$84.41\%$ & $-$& 6	&40	&-2 \\
			Right FP & $4$ &  $1327$  & $93.97\%$&  $93.97\%$ &$84.32\%$ & $-$& 30&	56&	28\\
			Left SG/AG & $4$ &  $859$  & $90.69\%$&  $90.8\%$ &$75.79\%$ & $-$& -50&	-56&	36 \\
			Right FP & $4$ &  $243$  & $67.08\%$&  $67.08\%$ &$43.21\%$ & $-$& 30&	64&	-4 \\
			Left SFG & $4$ & 	$202$  & $61.88\%$&  $61.88\%$ &$40.1\%$ & $-$& -18	&8	&52 \\
			Right SFG & $4$ & $122$  & $46.72\%$&  $46.72\%$ &$19.67\%$ & $-$&22&	10	&52  \\ 
			\rowcolor{lightgray}
			Right STG/PT/MTG & $3.2$ &$12540$  &   $89.82\%$& $89.86\%$ & $83.49\%$ &$<0.0001$ &60	&-10&	0 \\
			\rowcolor{lightgray}
			HG/PrG/T & &  &  &  & &  &  &  & \\
			STG/PT/MTG/HG &$4$ &$9533$  &   $99.16\%$ &  $99.16\%$ & $97.8\%$ &$-$& 60&	-10	&0 \\
			PrG & $4$ &  $485$  & $86.19\%$&  $86.19\%$ &$78.35\%$ & $-$& 52&	0&	48 \\
			T & $4$ &  $292$  & $72.6\%$&  $72.6\%$ &$53.77\%$ & $-$& 10&	-10&	8 \\
			\rowcolor{lightgray}
			Left STG/PT/MTG/&  $3.2$ &  $10833$  &     $88.4\%$&  $88.45\%$ & $80.41\%$ &$<0.0001$& -60 &	-12&	2 \\
			\rowcolor{lightgray}
			HG/IFG/T & &  &  &  & &  &  &  & \\
			HG/PT/MTG/STG &$4$ &$7894$  &   $98.99\%$ &  $98.99\%$ & $97.35\%$ &$-$ &-60&	-12&	2 \\
			IFG & $4$ &  $667$  & $88.01\%$&  $88.19\%$ &$74.06\%$ & $-$& -40&	14&	26 \\
			T & $4$ & $34$  & $26.47\%$&  $26.47\%$ &$17.65\%$ & $-$& -14&	-26&	-4 \\
			\rowcolor{lightgray}
			Right IC/CO & $3.2$ &  $408$  & $37.25\%$&  $37.25\%$ &$24.01\%$ &$0.0002$ & 38	&-2&	16 \\
			- & $4$ & $226$  & $67.26\%$&  $67.26\%$ &$43.36\%$ & $-$& 38&	-2&	16 \\ 
			\rowcolor{lightgray}
			Left PrG & $3.2$ &  $276$   &     $49.63\%$&  $49.64\%$  & $43.84\%$  &$0.002$&-52&	-6&	50 \\
			- & $4$ &  $192$  & $71.35\%$&  $71.35\%$ &$63.02\%$ & $-$& -52&	-6&	50 \\ 
			\rowcolor{lightgray}
			FM& $3.2$ &  $270$   &     $22.59\%$&  $22.59\%$  & $13.33\%$  &$0.002$&4&	50&	-14 \\
			- & $4$ & $128$  & $47.66\%$&  $47.66\%$ &$28.13\%$ & $-$& 4&	50&	-14 \\ 
			\rowcolor{lightgray}
			SFG& $3.2$ &  $187$   &     $6.95\%$&  $6.95\%$  & $0\%$  &$0.0123$&6&	52&	38 \\
			- & $4$ & $64$  & $20.31\%$&  $21.23\%$ &$0\%$ & $-$& 6&	52&	38 \\
			\rowcolor{lightgray}
			Left T& $3.2$ &  $176$   &     $1.14\%$&  $1.14\%$  & $0\%$  &$0.0157$&-14	&-14&	10 \\
			- & $4$ & $49$  & $4.08\%$&  $4.08\%$ &$0\%$ & $-$& -14&	-14	&10 \\
			\hline
		\end{tabular}
			\end{adjustbox}
\end{table}

\begin{table}
	\centering
	\caption{Auditory data: Clusters $S$ identified with threshold $\mathbf{|T|}>3.2$ and $\bar{\pi}(S)$ computed using Simes family of critical vectors with $\delta \in \{0, 9, 27\}$. The iterative method is applied. The size of the clusters $|S|$ and the voxel coordinates $(x,y,z)$ are reported for each cluster.}    
	\label{tab:AuditorySimesDelta}
		\begin{adjustbox}{max width=\textwidth}
		\begin{tabular}{@{}lrrrrrrr@{}}
			\hline
			Cluster & Size &\multicolumn{3}{c}{$\%$ active} & \multicolumn{3}{c}{voxel coordinates}     \\
			$S$ & $|S|$ & \multicolumn{3}{c}{$\bar{\pi}(S)$} &  x & y & z \\ 
			&  & $\delta=0$ &$\delta=9$ &$\delta=27$ &   &  &  \\ 
			\hline
			FP/CG/SFG/TOF/LO/LG/OFG/ITG/SG/AG &$40094$  &   $96.33\%$ &  $97.01\%$ & $97.31\%$ & -30	&-34&	-16 \\
			Right STG/PT/MTG/HG/PrG/T &$12540$  &   $89.09\%$& $90.56\%$ & $91.43\%$  &60	&-10&	0 \\
			Left STG/PT/MTG/HG/IFG/T&    $10833$  &     $87.33\%$&  $89.25\%$ & $90.1\%$ & -60 &	-12&	2 \\
			Right IC/CO &   $408$  & $36.03\%$&  $36.76\%$ &$33.57\%$ & 38	&-2&	16 \\
			Left PrG &  $276$   &     $49.28\%$&  $47.46\%$  & $42.03\%$  &-52&	-6&	50 \\
			FM&   $270$   &     $21.11\%$&  $21.11\%$  & $16.67\%$  &4&	50&	-14 \\
			SFG&   $187$   &     $6.42\%$&  $3.74\%$  & $0\%$  &6&	52&	38 \\
			Left T&  $176$   &     $1.14\%$&  $0\%$  & $0\%$  &-14	&-14&	10 \\
			\hline
		\end{tabular}
			\end{adjustbox}
\end{table}
\begin{table}
	\centering
	\caption{Auditory data: Clusters $S$ identified with threshold $\mathbf{|T|}>3.2$ and $\bar{\pi}(S)$ computed using AORC family of critical vectors with $\delta \in \{0, 9, 27\}$. The iterative method is applied. The size of the clusters $|S|$ and the voxel coordinates $(x,y,z)$ are reported for each cluster.}     
	\label{tab:AuditoryFinnerDelta}
		\begin{adjustbox}{max width=\textwidth}
		
		\begin{tabular}{@{}lrrrrrrr@{}}
			\hline
			Cluster & Size &\multicolumn{3}{c}{$\%$ active} & \multicolumn{3}{c}{voxel coordinates}     \\
			$S$ & $|S|$ & \multicolumn{3}{c}{$\bar{\pi}(S)$} &  x & y & z \\ 
			&  & $\delta=0$ &$\delta=9$ &$\delta=27$ &   &  &  \\ 
			\hline
			FP/CG/SFG/TOF/LO/LG/OFG/ITG/SG/AG &$40094$  &   $96.37\%$ &  $97.04\%$ & $97.33\%$ & -30	&-34&	-16 \\
			Right STG/PT/MTG/HG/PrG/T &$12540$  &   $89.15\%$& $90.64\%$ & $91.5\%$  &60	&-10&	0 \\
			Left STG/PT/MTG/HG/IFG/T&    $10833$  &     $87.42\%$&  $89.34\%$ & $90.16\%$ & -60 &	-12&	2 \\
			Right IC/CO &   $408$  & $36.28\%$&  $36.76\%$ &$33.58\%$ & 38	&-2&	16 \\
			Left PrG &  $276$   &     $49.28\%$&  $47.46\%$  & $42.03\%$  &-52&	-6&	50 \\
			FM&   $270$   &     $21.11\%$&  $21.11\%$  & $16.67\%$  &4&	50&	-14 \\
			SFG&   $187$   &     $6.42\%$&  $3.74\%$  & $0\%$  &6&	52&	38 \\
			Left T&  $176$   &     $1.14\%$&  $0\%$  & $0\%$  &-14	&-14&	10 \\
			\hline
		\end{tabular}
			\end{adjustbox}
\end{table}

\begin{table}
	\centering
	\caption{Auditory data: Clusters $S$ identified with threshold $\mathbf{|T|}>3.2$ and $\bar{\pi}(S)$ computed using family of critical vectors based on the Higher Criticism. The single-step method is applied. The size of the clusters $|S|$ and the voxel coordinates $(x,y,z)$ are reported for each cluster.}  
	\label{tab:AuditoryHCBeta}
		\begin{adjustbox}{max width=\textwidth}
		\begin{tabular}{@{}lrrrrr@{}}
			\hline
			Cluster & Size &\multicolumn{1}{c}{$\%$ active} & \multicolumn{3}{c}{voxel coordinates} \\
			$S$ & $|S|$ & \multicolumn{1}{c}{$\bar{\pi}(S)$} &  x & y & z \\ 
			&  & Higher criticism   &  & &  \\ 
			\hline
			FP/CG/SFG/TOF/LO/LG/OFG/ITG/SG/AG &$40094$  &   $95.66\%$ &   -30	&-34&	-16 \\
			Right STG/PT/MTG/HG/PrG/T &$12540$  &   $86.17\%$&  60	&-10&	0 \\
			Left STG/PT/MTG/HG/IFG/T&    $10833$  &     $83.99\%$&    -60 &	-12&	2 \\
			Right IC/CO &   $408$  & $0\%$&   38	&-2&	16 \\
			Left PrG &  $276$   &     $24.64\%$&   -52&	-6&	50 \\
			FM&   $270$   &     $0\%$&   4&	50&	-14 \\
			SFG&   $187$   &     $0\%$&   6&	52&	38 \\
			Left T&  $176$   &     $0\%$&   -14	&-14&	10 \\
			\hline
		\end{tabular}
			\end{adjustbox}
\end{table}

\section{Rhyme data analysis}\label{rhyme_plus}

Table \ref{tab:Food1} includes the results applying the single-step method presented in Section \ref{methods} imposing $\delta = 27$, following the structure of Table \ref{tab:Food} of Subsection \ref{food-data}. The performance having $\delta \in \{0, 1, 9\}$ is proposed in Table \ref{tab:FoodSimesDelta} using the Simes family of confidence bounds, and in Table \ref{tab:FoodFinnerDelta} using the AORC family. Table \ref{tab:hc_it} shows the results using the family of critical vectors based on the Higher Criticism. In all the analysis, the $\alpha$ level equals $0.05$. \angela{Finally, Table \ref{tab:FoodTFCE} shows the lower bounds for the TDP as before but considering clusters computed by threshold-free cluster enhancement (TFCE) method \citep{smith2009threshold} using p-value threshold equals $0.05$. We found activity in Lingual Gyrus (LG), Occipital Pole (OP), Putamen (P), Superior Frontal Gyrus (SFG), Frontal Pole (FP), Insular Cortex (I), Occipital Fusiform Gyrus (OFG), Lateral Occipital Cortex (LO), Precentral Gyrus (PrG), Post Central Gyrus (PCG), and Paracingulate Gyrus (PG).}

\begin{table}
	\centering
	\caption{Rhyme data: Clusters $S$ identified with threshold $\mathbf{|T|}>3.2$ and active proportion percentage $\bar{\pi}(S)$ using Simes and AORC families ($\delta = 27$) and parametric ARI, ``drill down'' clusters at $\mathbf{|T|}>4$. The single-step method is applied. The size of the clusters $|S|$ and the voxel coordinates $(x,y,z)$ are reported for each cluster.}       
	\label{tab:Food1}
		\begin{adjustbox}{max width=\textwidth}
		\begin{tabular}{@{}rlrrrrcccc@{}}
			\hline
			\multicolumn{1}{c}{Cluster} &\multicolumn{1}{c}{Threshold} & \multicolumn{1}{c}{Size} &\multicolumn{3}{c}{$\%$ active}& \multicolumn{1}{c}{RFT} & \multicolumn{3}{c}{Voxel}     \\
			&  &  && & &\multicolumn{1}{c}{P-Values}& \multicolumn{3}{c}{Coordinates}     \\
			\multicolumn{1}{c}{$S$}&\multicolumn{1}{c}{$t$}  &\multicolumn{1}{c}{$|S|$} & \multicolumn{3}{c}{$\bar{\pi}(S)$}&\multicolumn{1}{c}{$p_{FWER}$}  &   \multicolumn{1}{c}{x} & \multicolumn{1}{c}{y} & z \\ 
			&  &&Perm &Perm & Parametric & &   &  &  \\ 
			&&  &Simes \refstepcounter{equation}(\theequation)\label{eq:tblSimes1}&AORC \refstepcounter{equation}(\theequation)\label{eq:tblAORC1} &Simes \refstepcounter{equation}(\theequation)\label{eq:tblpar1}  &  &     &  &\\
			\hline
			\rowcolor{lightgray}
			LOC/LG/OFG/PG/SFG & $3.2$ & $34115$  &  $87.38\%$ & $87.85\%$ &  $38.16\%$ & $<0.001$ & $4$ &  $12$ &  $48$ \\
			\rowcolor{lightgray}
			FOC/P/IFG/IC/CG & &  &    &   &  & &  &  &   \\
			LOC/LG/OFG & $4$ &$11045$ &  $90.82\%$ & $91.09\%$ &  $42.01\%$& $-$ & $-6$ &  $-56$ &  $-12$ \\
			FOC/P/IFG/IC &$4$ & $6930$   & $85.38\%$ & $85.81\%$ &  $29.32\%$& $-$ & $-42$ &  $14$ &  $-6$ \\
			PG/SFG/CG & $4$ &  $2100$  & $56.95\%$ & $57.67\%$ & $18.05\%$& $-$ & $4$ & $12$ &  $48$ \\
			Left P & $4$ &  $38$  & $2.63\%$ & $2.63\%$ & $2.63\%$& $-$ & $-32$ &  $-18$ &  $-8$ \\
			\rowcolor{lightgray}
			Left SPL/PCG & $3.2$ & $1546$ & $1.49\%$ & $1.75\%$ &  $0\%$ & $<0.001$ & $-24$ &  $-62$ &  $44$ \\
			\hline
		\end{tabular}
			\end{adjustbox}
\end{table}

\begin{table}
	\centering
	\caption{Rhyme data: Clusters $S$ identified with threshold $\mathbf{|T|}>3.2$ and $\bar{\pi}(S)$ computed using Simes family of critical vectors with $\delta \in \{0, 1, 9\}$. The iterative method is applied. The size of the clusters $|S|$ and the voxel coordinates $(x,y,z)$ are reported for each cluster. }
	\label{tab:FoodSimesDelta}
		\begin{adjustbox}{max width=\textwidth}
		\begin{tabular}{@{}lrrrrrrr@{}}
			\hline
			Cluster & Size &\multicolumn{3}{c}{$\%$ active} & \multicolumn{3}{c}{voxel coordinates}     \\
			$S$ & $|S|$ & \multicolumn{3}{c}{$\bar{\pi}(S)$} &  x & y & z \\ 
			&  & $\delta=0$ &$\delta=1$ &$\delta=9$ &   &  &  \\ 
			\hline
			LOC/LG/OFG/PG/SFG/FOC/P/IFG/IC/CG & $11045$ &   $67.48\%$& $84.23\%$ & $87.27\%$ & $4$ &  $12$ &  $48$ \\
			Left SPL/PCG & $1546$  & $0\%$ & $1.55\%$ & $2.52\%$ & $-24$ &  $-62$ &  $44$ \\
		\end{tabular}
			\end{adjustbox}
\end{table}
\begin{table}
	\centering
	\caption{Rhyme data: Clusters $S$ identified with threshold $\mathbf{|T|}>3.2$ and $\bar{\pi}(S)$ computed using AORC family of critical vectors with $\delta \in \{0, 1, 9\}$. The iterative method is applied. The size of the clusters $|S|$ and the voxel coordinates $(x,y,z)$ are reported for each cluster.} 
	\label{tab:FoodFinnerDelta}
		\begin{adjustbox}{max width=\textwidth}
		\begin{tabular}{@{}lrrrrrrr@{}}
			\hline
			Cluster & Size &\multicolumn{3}{c}{$\%$ active} & \multicolumn{3}{c}{voxel coordinates}     \\
			$S$ & $|S|$ & \multicolumn{3}{c}{$\bar{\pi}(S)$} &  x & y & z \\ 
			&  & $\delta=0$ &$\delta=1$ &$\delta=9$ &   &  &  \\ 
			\hline
			LOC/LG/OFG/PG/SFG/FOC/P/IFG/IC/CG & $3331$   &   $68.67\%$& $84.66\%$ & $87.65\%$ & $4$ &  $12$ &  $48$ \\
			Left SPL/PCG & $1546$  & $0\%$ & $1.68\%$ & $2.72\%$ & $-24$ &  $-62$ &  $44$ \\
		\end{tabular}
			\end{adjustbox}
\end{table}

\begin{table}
	\centering
	\caption{Rhyme data: Clusters $S$ identified with threshold $\mathbf{|T|}>3.2$ and $\bar{\pi}(S)$ computed using family of critical vectors based on the Higher Criticism. The single-step method is applied. The size of the clusters $|S|$ and the voxel coordinates $(x,y,z)$ are reported for each cluster.}
	\label{tab:hc_it}
		\begin{adjustbox}{max width=\textwidth}
		\begin{tabular}{@{}lrrrrr@{}}
			\hline
			Cluster & Size &\multicolumn{1}{c}{$\%$ active} & \multicolumn{3}{c}{voxel coordinates}     \\
			$S$ & $|S|$ & \multicolumn{1}{c}{$\bar{\pi}(S)$} &  x & y & z \\ 
			&  & Higher Criticism &   & &  \\ 
			\hline
			LOC/LG/OFG/PG/SFG/FOC/P/IFG/IC/CG & $3331$   &   $84.1\%$&   $4$ &  $12$ &  $48$ \\
			Left SPL/PCG & $1546$  & $0\%$ &  $-24$ &  $-62$ &  $44$ \\
		\end{tabular}
			\end{adjustbox}
\end{table}

\begin{table}
	\centering
	\caption{\angela{Rhyme data: Clusters $S$ identified by threshold-free cluster enhancement (TFCE) method \cite{smith2009threshold} and Active Proportion Percentage $\bar{\pi}(S)$ using Simes and AORC families ($\delta = 27$) and parametric ARI. The size of the clusters $|S|$ is reported for each cluster.}}    
	\label{tab:FoodTFCE}
		\begin{adjustbox}{max width=\textwidth}
		\begin{tabular}{@{}rrrrr@{}}
			\hline
			\multicolumn{1}{c}{Cluster} & \multicolumn{1}{c}{Size} &\multicolumn{3}{c}{$\%$ active}    \\
			\multicolumn{1}{c}{$S$} &\multicolumn{1}{c}{$|S|$} & \multicolumn{3}{c}{$\bar{\pi}(S)$} \\ 
			& &Perm &Perm & Parametric \\ 
			&&  Simes \refstepcounter{equation}(\theequation)\label{eq:tblSimes11TFCE}&AORC \refstepcounter{equation}(\theequation)\label{eq:tblAORC11TFCE} &Simes \refstepcounter{equation}(\theequation)\label{eq:tblpar11TFCE}  \\
			\hline
			LG/OP/P/SFG/FP/I/OFG/LO/PrG/PCG/PG  & $73401$  &  $44.76\%$ & $45.57\%$ &  $38.78\%$ \\
		\end{tabular}
			\end{adjustbox}
\end{table}

\section{Word object analysis}

We analyze the dataset provided by \cite{Duncan}. It consists of $48$ subjects looking $4$ different visual stimuli: written words, pictures of objects, scrambled pictures of the same objects, and consonant letter strings. It is a block design where each block contains $16$ stimuli from a single category using a one-back task, i.e., two runs are performed. Therefore, in this case, a third-level analysis was carried out. Table \ref{tab:WO} reports the results regarding the contrast of the activation difference between word and consonant string stimuli, considering the single-step method $\alpha$ equals $0.05$ and $\delta =1$. We found activation in Intracalcarine Cortex (IC), Lingual Gyrus (LG), Precentral Gyrus (PrG), Cuneal Cortex (CC), Planum Temporale (PT), Supramarginal Gyrus (SG), Amygdala (A), Superior Temporal Gyrus (STG), Insular Cortex (I), Lateral Occipital Cortex (LO), Middle Frontal Gyrus (MFG), Precuneous Cortex (PrC), Cingulate Gyrus (CG), Accumbens (Ac), Central Opercular Cortex (CO), Thalamus (T), and Superior Frontal Gyrus (SFG).

\begin{figure}
	\centering
	\includegraphics[width=.8\textwidth]{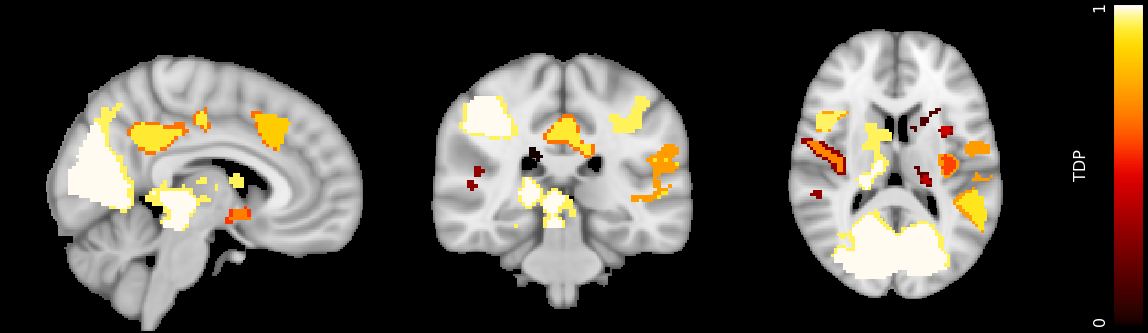}
	\caption{Word-Object data: True Discovery Proportion map using the Simes family of critical vectors with $\delta = 1$. Colors express the True Discovery Proportion for clusters corresponding to a threshold of $3.2$ and ``drilled'' down at $4$.}
	\label{fig:WO_tdp}
\end{figure}

Figure \ref{fig:WO_tdp} shows the TDP as a cluster brain map regarding the results using the Simes family confidence bound.

If you are interested in analyzing other possible contrasts, e.g., words versus scrambled pictures, you can find the full dataset in \angela{\url{https://github.com/angeella/fMRIdata}} \citep{fMRIdata}.

\begin{table}
	\centering
	\caption{Word-Object data: Clusters $S$ identified with threshold $\mathbf{|T|}>3.2$ and active proportion percentage $\bar{\pi}(S)$ using Simes and AORC families ($\delta = 1$) and parametric ARI, and ``drill down'' clusters at $\mathbf{|T|}>4$. The single-step method is applied. The size of the cluster $|S|$ and the voxel coordinates $(x,y,z)$ are reported for each cluster.}
	\label{tab:WO}
		\begin{adjustbox}{max width=\textwidth}
		\begin{tabular}{@{}rlrrrrrrrr@{}}
			\hline
			\multicolumn{1}{c}{Cluster} &\multicolumn{1}{c}{Threshold} & \multicolumn{1}{c}{Size} &\multicolumn{3}{c}{$\%$ active}& \multicolumn{1}{c}{RFT} & \multicolumn{3}{c}{Voxel}     \\
			&  &  && & &\multicolumn{1}{c}{P-Values}& \multicolumn{3}{c}{Coordinates}     \\
			\multicolumn{1}{c}{$S$}&\multicolumn{1}{c}{$t$}  &\multicolumn{1}{c}{$|S|$} & \multicolumn{3}{c}{$\bar{\pi}(S)$}&\multicolumn{1}{c}{$p_{FWER}$}  &   \multicolumn{1}{c}{x} & \multicolumn{1}{c}{y} & z \\ 
			&  &&Perm &Perm & Parametric & &   &  &  \\ 
			&&  &Simes &AORC &Simes  &  &     &  &\\
			\hline
			\rowcolor{lightgray}
			IC/LG/PrC/CC & $3.2$ & $17431$ & $94.52\%$& $94.56\%$ & $84.31\%$ & $<0.0001$ &-2 &	-78	&10 \\
			IC/LG/PrC & $4$ & $13469$ & $99.35\%$& $99.35\%$ & $97.95\%$ & $-$ &-2	&-78&	10 \\
			\rowcolor{lightgray}
			Left PT/SG/A/STG/I/LO & $3.2$ & $3516$ & $73.23\%$& $73.43\%$ & $46.62\%$ & $<0.0001$ &-24&	-14	&-12 \\
			PT/SG & $4$ & $888$ & $90.31\%$& $90.32\%$ & $69.93\%$ & $-$ &-56&	-44&	14 \\
			A & $4$ & $382$ & $78.27\%$& $78.27\%$ & $54.19\%$ & $-$ &-24&	-14	&-12 \\
			STG & $4$ & $289$ & $74.74\%$& $74.74\%$ & $56.75\%$ & $-$ &-52&	-12&	-6 \\
			I & $4$ & $117$ & $56.41\%$& $56.41\%$ & $31.62\%$ & $-$ &-34	&-16&	16\\
			LO & $4$ & $44$ & $4.55\%$& $4.55\%$ & $0\%$ & $-$ &-42&	-64	&48\\
			\rowcolor{lightgray}
			Right MFT/PrG & $3.2$ & $2217$ & $74.92\%$& $74.97\%$ & $62.65\%$ & $<0.0001$ &26&	6&	48 \\
			- & $4$ & $1658$ & $94.75\%$& $94.75\%$ & $83.78\%$ & $-$ &26	&6&	48\\
			\rowcolor{lightgray}
			CG & $3.2$ & $1640$ & $64.57\%$& $64.57\%$ & $52.68\%$ & $<0.0001$ &-4 &	-44	&32 \\
			- & $4$ & $1101$ & $92.1\%$& $92.1\%$ & $78.47\%$ & $-$ &-4	&-44&	32 \\
			\rowcolor{lightgray}
			Right STG/A/Ac & $3.2$ & $1354$ & $55.1\%$& $55.1\%$ & $38.47\%$ & $<0.0001$ &24&	-14&	-16 \\
			STG & $4$ & $345$ & $77.97\%$& $77.97\%$ & $54.2\%$ & $-$ &58&	-10&	-8\\
			A & $4$ & $258$ & $66.67\%$& $66.67\%$ & $41.47\%$ & $-$ &24&	-14	&-16\\
			Ac & $4$ & $168$ & $66.67\%$& $66.67\%$ & $45.23\%$ & $-$ &-4&	8&	-10\\
			\rowcolor{lightgray}
			Right CO & $3.2$ & $792$ & $29.17\%$& $29.17\%$ & $7.2\%$ & $<0.0001$ &50&	-6&	8 \\
			- & $4$ & $270$ & $67.78\%$& $67.78\%$ & $21.11\%$ & $-$ &50&	-6&	8 \\
			\rowcolor{lightgray}
			PG/CG & $3.2$ & $637$ & $63.42\%$& $63.42\%$ & $53.06\%$ & $<0.0001$ &4&	24	&36 \\
			- & $4$ & $480$ & $84.17\%$& $84.17\%$ & $70.42\%$ & $-$ &4&	24&	36\\
			\rowcolor{lightgray}
			Right LO & $3.2$ & $603$ & $41.79\%$& $41.79\%$ & $2.338\%$ & $<0.0001$ &46&	-64	&40 \\
			- & $4$ & $331$ & $73.72\%$& $73.72\%$ & $42.6\%$ & $-$ &46&	-64&	40\\
			\rowcolor{lightgray}
			Left SFG & $3.2$ & $449$ & $41.2\%$& $41.2\%$ & $28.06\%$ & $<0.0001$ &-24&	4&	46 \\
			- & $4$ & $266$ & $69.55\%$& $69.55\%$ & $47.37\%$ & $-$ &-24&	4&	46 \\
			\rowcolor{lightgray}
			Right T & $3.2$ & $197$ & $17.26\%$& $17.26\%$ & $8.63\%$ & $0.0003$ &-20&	-26	&6 \\
			- & $4$ & $86$ & $39.53\%$& $39.54\%$ & $19.77\%$ & $-$ &-20&	-26&	6\\
			\rowcolor{lightgray}
			- & $3.2$ & $191$ & $1.57\%$& $1.57\%$ & $0\%$ & $0.0004$ &24&	-40	&20 \\
			- & $4$ & $47$ & $2.13\%$& $2.13\%$ & $0\%$ & $-$ &24	&-40&	20\\
			\rowcolor{lightgray}
			Left I & $3.2$ & $188$ & $13.82\%$& $13.83\%$ & $5.85\%$ & $0.0005$ &-28&	16&	4 \\
			- & $4$ & $62$ & $41.94\%$& $41.94\%$ & $17.74\%$ & $-$ &-28&	16&	4\\
			\rowcolor{lightgray}
			IC & $3.2$ & $58$ & $25.86\%$& $25.86\%$ & $18.97\%$ & $0.084$ &-32&	6&	10 \\
			\hline
		\end{tabular}
			\end{adjustbox}
\end{table}

\section{Iterative approach}\label{it_plus}
In this section, the iterative version, defined in Theorem \ref{iterative_thm1}, is examined following the simulation analysis proposed in Section \ref{Simulation}. The method is applied directly on $S_{\pi_1}$ the set of true discoveries, therefore $\pi(S_{\pi_1}) = 1$.

\begin{figure}
	\centering
	\includegraphics[width=.8\textwidth]{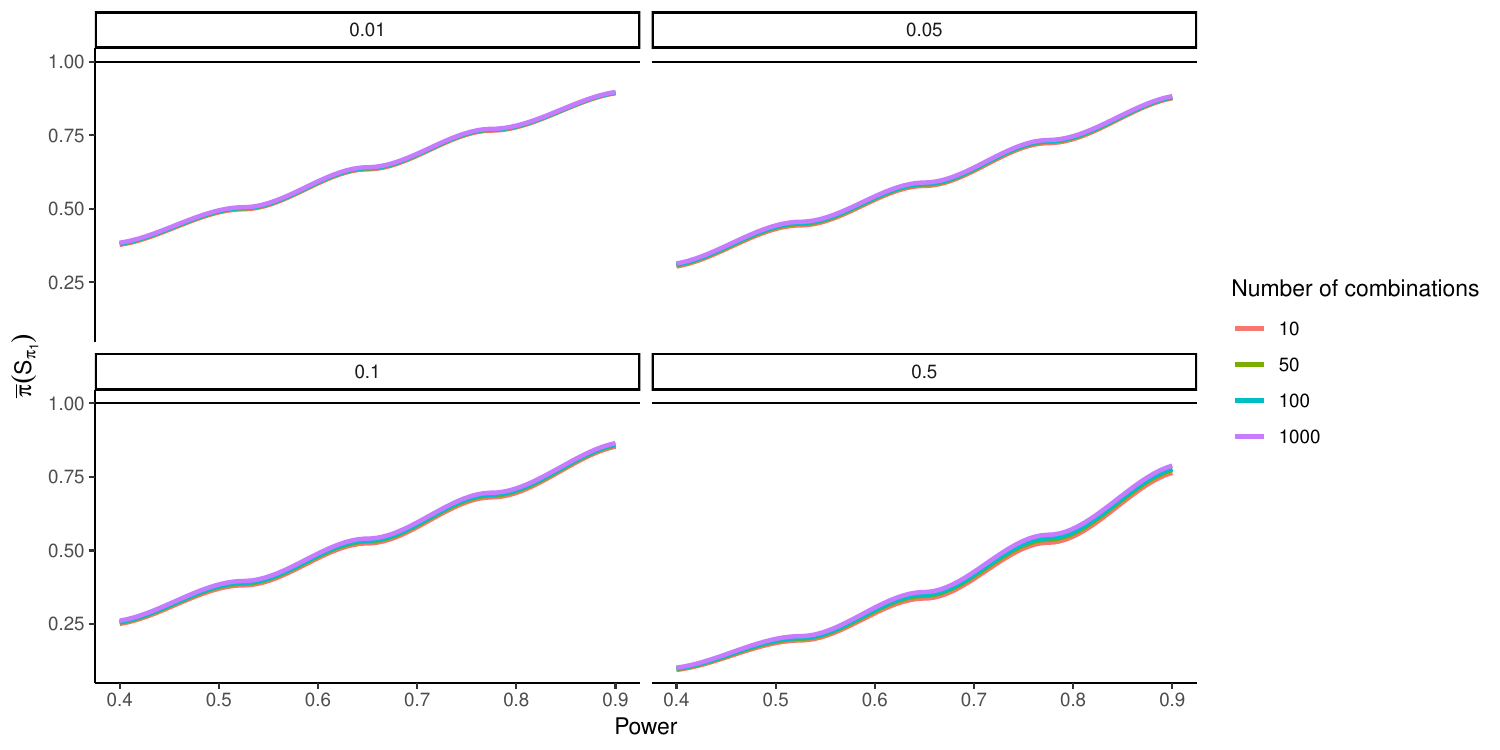}
	\caption{Simulated true discovery lower bounds for $S_{\pi_1}$ over different values of $\theta$ and power using the approximated iterative version with $10$, $50$, $100$, and, $1000$ random combinations. The solid black line represents the true discovery proportion $\pi(S_{\pi_1}) = 1$.}
	\label{fig:ncomb1}
\end{figure}

\begin{figure}
	\centering
	\includegraphics[width=.6\textwidth]{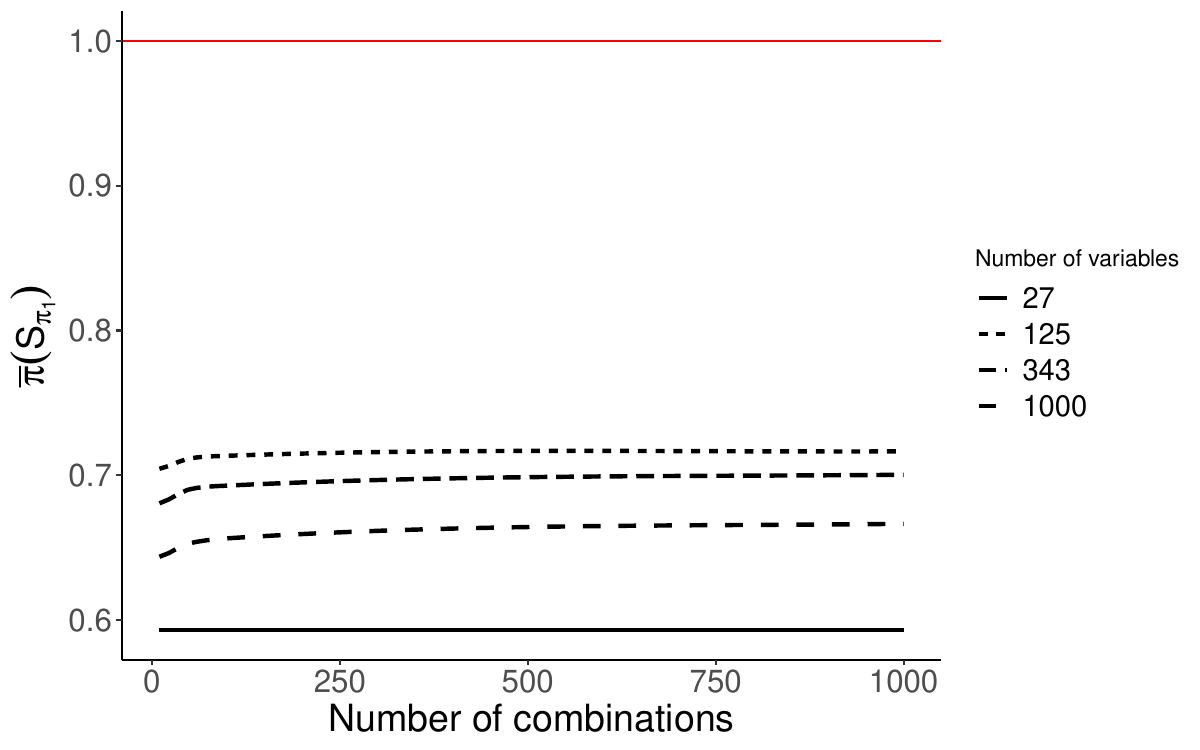}
	\caption{Simulated true discovery lower bounds for $S_{\pi_1}$ over different values of $m$ variables using the approximated iterative version with $10, \dots, 100$ random combinations. The solid red line represents the true discovery proportion $\pi(S_{\pi_1}) = 1$.}
	\label{fig:ncomb_m}
\end{figure}

Figure \ref{fig:ncomb1} illustrates the behavior of the approximated iterative method using a different number of combinations. The approximation version becomes exact when the number of combinations goes to infinity. However, as we can see in Figure \ref{fig:ncomb1}, the results using only $10$ combinations are nearly equal to the results using $1000$ combinations. In addition, looking at Figure \ref{fig:ncomb_m}, we can see that the method is robust if a different number of variables are considered, i.e., the lines in Figure \ref{fig:ncomb_m} are below $1$ (true discovery proportion). \angela{ In this case, we fix $\theta = 0.2$.} 

\section{Proof of Theorem 1}\label{thm_proof}
\begin{proof}
	In Lemma $6$ by \cite{GoemanSolari}, define $l_{i:n} = l_i$ for every $i \ge 1$ and $n \ge 0$. This lemma then implies that:
	\begin{align}
		\max_{1 \le u \le |S|} 1 - u + |\{i \in S: p_i \le l_u \}|
	\end{align}
	are valid simultaneous bounds, as was to be shown.
\end{proof}

\section{Proof of Theorem 2}\label{thm_proof_perm}

We do not report a formal proof of Theorem $2$ since you can directly refer to \cite[p. 643]{Jesse}. Indeed, the relationship between the definition of the $\lambda_{\alpha}$ calibration parameter and the power of the method is evident, i.e., a large value of $\lambda_{\alpha}$ leads to more power. We mainly have rephrased \cite{Jesse}'s theorem based on the concept of upper bound for the false discovery proportion in terms of the $\lambda_{\alpha}$ calibration parameter. To sum up, Theorem $2$ gets an improvement of $\lambda_\alpha$ which gives an improved critical vector, and then using Theorem $1$ we get an improved lower $(1-\alpha)$ confidence bound of $a(S)$ simultaneously for all $S \subseteq B$.

\section{Validating Permutation-based ARI}\label{eklund_plus}

\begin{figure}
	\centering
	\includegraphics[width=.8\textwidth]{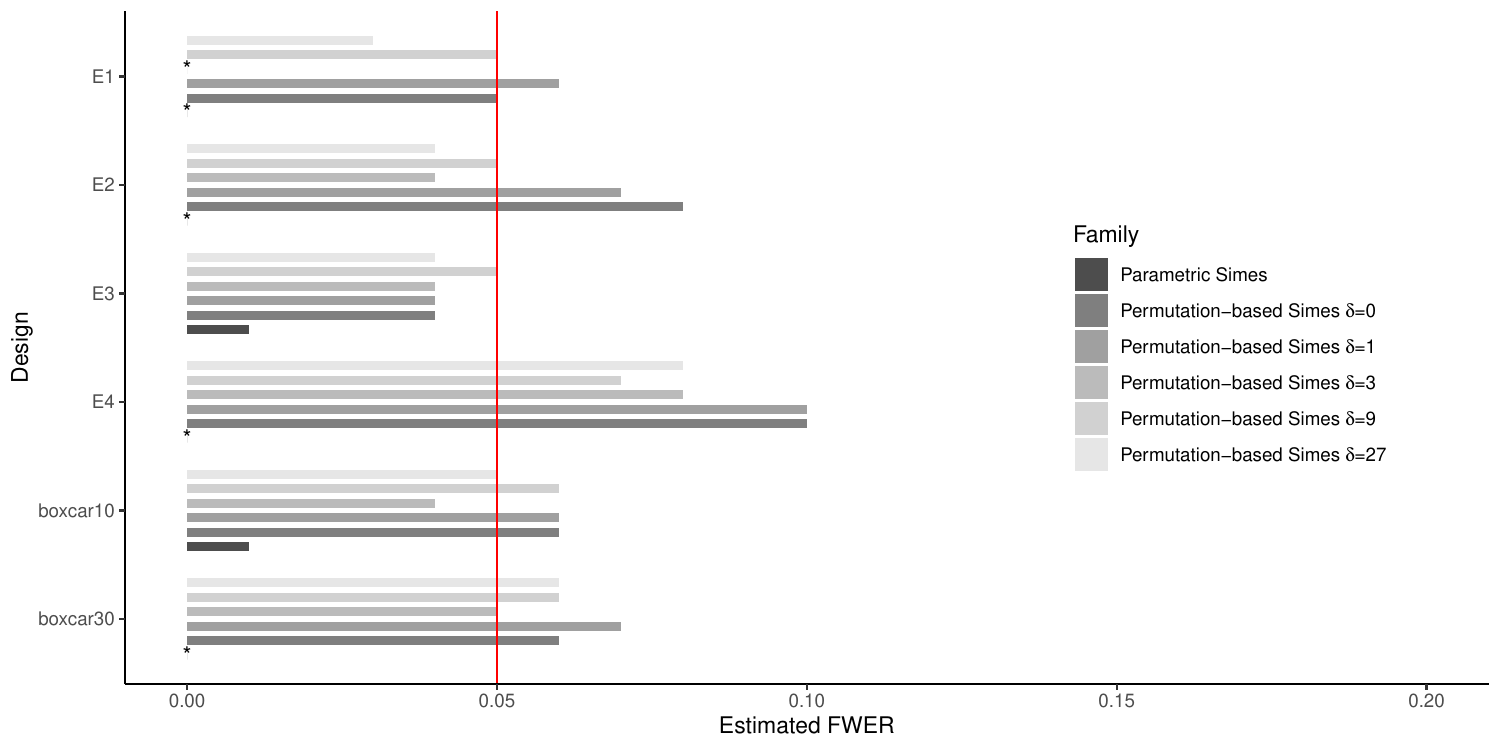}
	\caption{Estimated FWER considering six different first level designs, i.e., two-block activity paradigms: boxcar10 (10-s on-off), boxcar30 (30-s on-off), and four event activity paradigms, i.e., E1 (single event of 2-s activation, 6-s rest), E2 (single event 1- to 4-s activation, 3- to 6-s rest, randomized), E3 (13 events of 3–6 s for each task), and E4 (13 events of 3–6 s for each task, randomized), and six different methods to compute the TDP's lower bound (parametric Simes and permutation-based Simes considering five different values of the shift parameter, i.e., $\delta \in \{0, 1, 3, 9, 27\}$). The solid red lines represents \angela{the estimated }nominal FWER equals $0.05$, \angelaa{while the star symbols describe estimated FWER equals $0$.}}
	\label{oulu_os}
\end{figure}

We propose here the results of performing the one-sample t-tests instead of the two-sample t-tests in the Oulu data\angela{set}  \citep{Eklund} following the same procedure of Section \ref{validiting-and-power-of-the-permutation-based-ari-inference}. Figure \ref{oulu_os} is structured as Figure \ref{Oulu} presented in Section \ref{validiting-and-power-of-the-permutation-based-ari-inference}. Again, we can note how the parametric-based ARI returns in most cases an estimated FWER equals $0$, while the permutation-based ARI gain power controlling the FWER at the same time.

\section{Simulation study}\label{equi-corr}
We simulate data considering the simple following model:
\begin{align*}
	\angela{D_{ij} = \mu_i + \epsilon_{ij}^\star}
\end{align*}
where \angela{$\mathbf{D}_j \in \mathbb{R}^m$}, with $j = 1, \dots, J$, $J$ is the number of independent observations \angela{(i.e., subjects)} and $m$ is the total number of \angela{ voxels}. The noise \angela{$\epsilon_j^\star \in \mathbb{R}^m$} follows the multivariate normal distribution with mean $0$ and equi-correlation variance structure, i.e., $\epsilon_j^\star \sim \mathcal{N}(0,\Sigma_{\rho^2})$, where $\rho$ is the level of equi-correlation between pairs of \angela{voxels}. The signal $\boldsymbol{\mu}$ is computed considering the difference in means having power of the one-sample t-test equals $0.8$, \angela{i.e., $\boldsymbol{\mu} = (z_{1-\alpha/2} + z_{1-\beta})/\sqrt{J}$, where $\alpha  = 0.05$ is the significance level, $\beta = 0.8$ is the power level, and $z_{a}$ is the quantiles of the standard normal distribution at level $a$}. The signal $\boldsymbol{\mu}$ is equal to $0$ under the null hypothesis. 

\begin{figure}
	\centering
	\includegraphics[width=.8\textwidth]{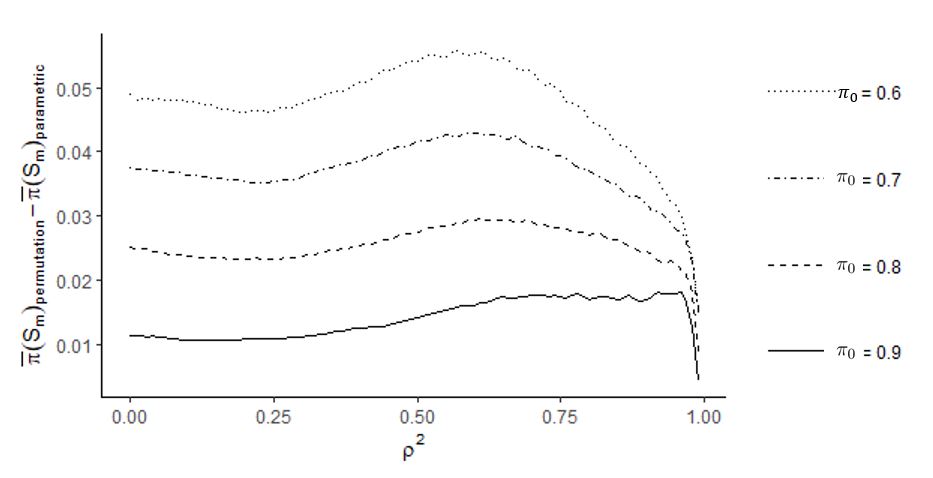}
	\caption{Difference of lower bounds for the true discoveries proportion considering the permutation $\bar{\pi}(S_m)_{\text{permutation}}$ and parametric $\bar{\pi}(S_m)_{\text{parametric}}$ methods using simulated data and considering the full set of hypotheses $S_m$ over different values of $\rho^2 \in \{0,0.01, \dots, 0.99,1\}$ and $\pi_0 \in \{0.6, 0.7, 0.8, 0.9 \}$.}
	\label{sim:pi_equi}
\end{figure}

First of all, we want to understand how the improvement of the nonparametric TDP lower bound changes concerning $\rho$ and the proportion of null hypotheses $\pi_0$. Let $J = 50$, $m = 1000$, $\rho^2 \in \{0,0.01, \dots, 0.99,1\}$ and $\pi_0 \in \{0.6, 0.7, 0.8, 0.9\}$, we simulate data $1000$ times and the mean of $\bar{\pi}(S_m)$ over simulation is represented. However, high values of $\rho^2$ are unrealistic in real applications. \angela{For example, the mean correlation across $10,000$ randomly sampled voxels equals $0.25$ in the case of Rhyme data.} The Simes family of confidence bound without shift is taken into account to compare with the parametric approach directly. Having no prior knowledge about the structure of the set of hypotheses to analyze, we consider the full set of hypotheses, i.e., $S_m$. Figure \ref{sim:pi_equi} shows the difference of $\bar{\pi}(S_m)$ computed using the permutation and parametric methods over the $\rho^2$ and $\pi_0$ values. As expected, the permutation approach gets some power with respect to the parametric one in the case of correlation between pairs of variables. It can handle any type of dependence structure of the $p$-values.

\begin{figure}
	\centering
	\includegraphics[width=.8\textwidth]{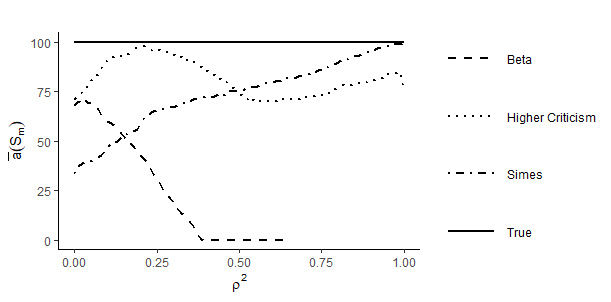}
	\caption{Simulated True Discovery lower bound over $S_m$ and different values of $\rho^2 \in \{0,0.01, \dots, 0.99,1\}$ using the Higher Criticism (dotted line), Beta (dashed line) and Simes critical vectors (dotted dashed line). The solid line represents the number of true discoveries which equals $100$ considering $1000$ variables and the proportion of null hypotheses $\pi_0 = 0.9$.}
	\label{sim:hc_beta_equi}
\end{figure}

Secondly, we want to examine why certain families of critical curves do not provide good results in Section \ref{applications}. The Higher Criticism critical vector \eqref{hc}, the Beta critical vector \eqref{beta} and the Simes critical vector \eqref{eq1} are then used to computed $\bar{a}(S_m)$ using simulated data with $\pi_0= 0.9$, $m = 1000$ and $J = 50$. As previously, we repeat the simulations $1000$ times for each framework, and the mean value of $\bar{a}(S_m)$ is computed. Figure \ref{sim:hc_beta_equi} shows the behaviour of these three families of critical vectors respect to $\rho^2 \in \{0,0.01,\dots, 0.99,1\}$. In Section \ref{lambdaCal}, we said that the Higher Criticism and Beta families could be problematic in the case of a strong correlation between tests. As expected, the Beta critical vector does not work in the case of a strong correlation between variables. This is due also to computational numerical difficulties, i.e., when the dashed line in Figure \ref{sim:hc_beta_equi} disappears. The Higher Criticism family seems to work, but it loses power with an increase in correlation.

\begin{figure}
	\centering
	\includegraphics[width=.8\textwidth]{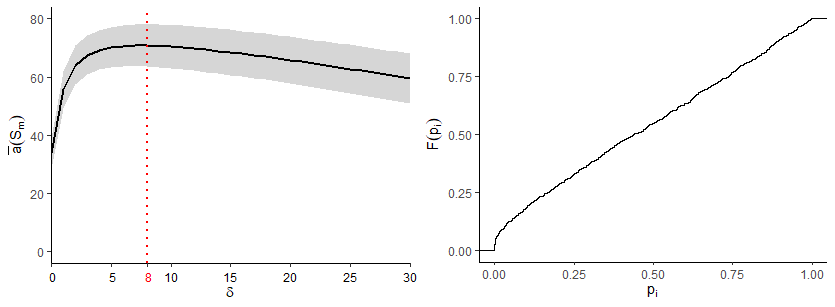}
	\caption{ Left side: True Discovery Lower Bound using simulated data. The full set of hypotheses $S_m$ is considered over different values of $\delta$. Right side: Empirical Cumulative Density Function of observed raw $p$-values, i.e, $F(p_i)$.}
	\label{sim:pv_equi}
\end{figure}

Thirdly, we want to analyze how the Simes family of critical curves \eqref{eq1} works if anti-conservative $p$-values distribution is considered. Let $J=50$, $m=1000$, $\rho = 0$ and $\pi_0=0.9$, we compute $\bar{a}(S_m)$ for each $1000$ simulations, and once again the mean over simulations is reported. Figure \ref{sim:pv_equi} shows $\bar{a}(S_m)$ considering the Simes family using $\delta \in \{0, \dots, 30\}$. We can note that the shifted version works well in the case of anti-conservative $p$-values if the corrected value for the tuning parameter $\delta$ is chosen, described by the red dotted line, i.e., $\delta = 8$. 

\begin{figure}
	\centering
	\includegraphics[width=.7\textwidth]{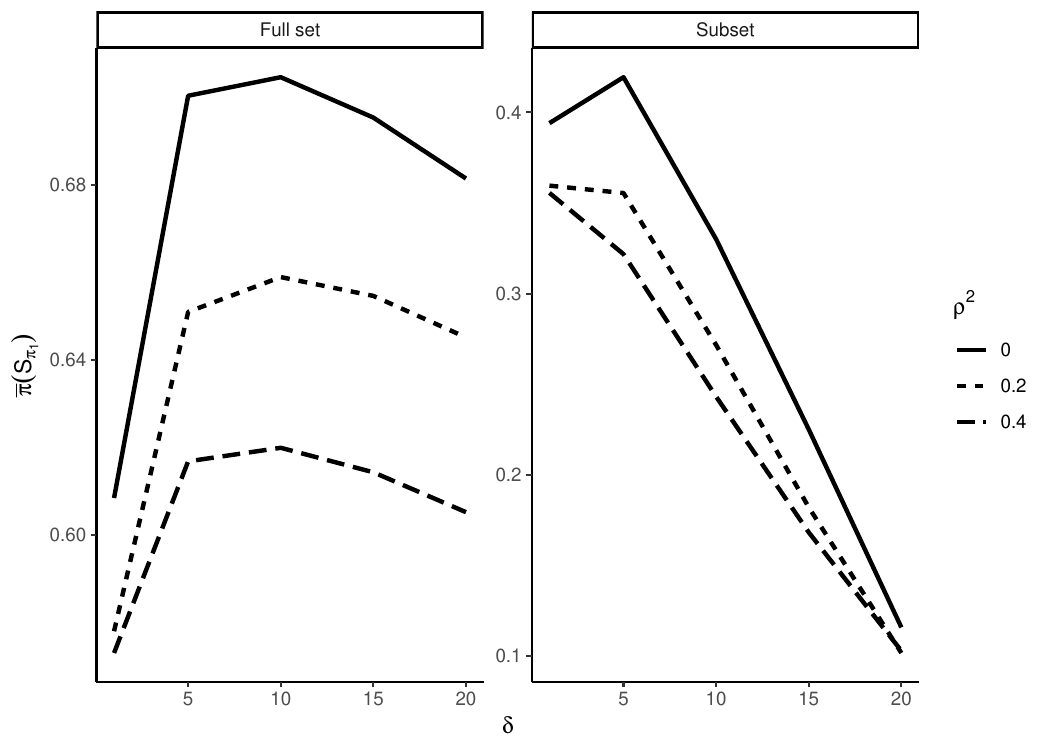}
	\caption{Lower Bounds for the True Discovery Proportion using simulated data the level of equi-correlation $\rho^2 \in \{0, 0.2, 0.4\}$. In the left figure, the full set of hypotheses is considered, while in the right figure a random sample of $40$ hypotheses is analyzed. The critical vectors based on the Simes family with $\delta \in \{0, 5, 10, 15, 20\}$ are used in both situations.}
	\label{sim:deltas_equi}
\end{figure}

Therefore, we explore how the Simes family of critical curves \eqref{eq1} works with different values of $\rho$ and $S$ size. The left part of Figure \ref{sim:deltas_equi} shows the mean of the lower bounds for the TDP considering the full set of hypotheses, i.e., $S_m$,  and $\delta \in \{0,5,10,15,20\}$ over $1000$ simulations. We can note that in almost all scenarios, the shifted version outperforms the unshifted ones. The difference gets smaller if $\rho^2$ increases. However, the situation changes if we compute the TDP for a smaller set of hypotheses than $S_m$ as shown in the right part of Figure \ref{sim:deltas_equi}. In this case, we randomly sample 40 hypotheses from the false null ones, i.e., $S_{40}$.

\begin{figure}
	\centering
	\includegraphics[width=\textwidth]{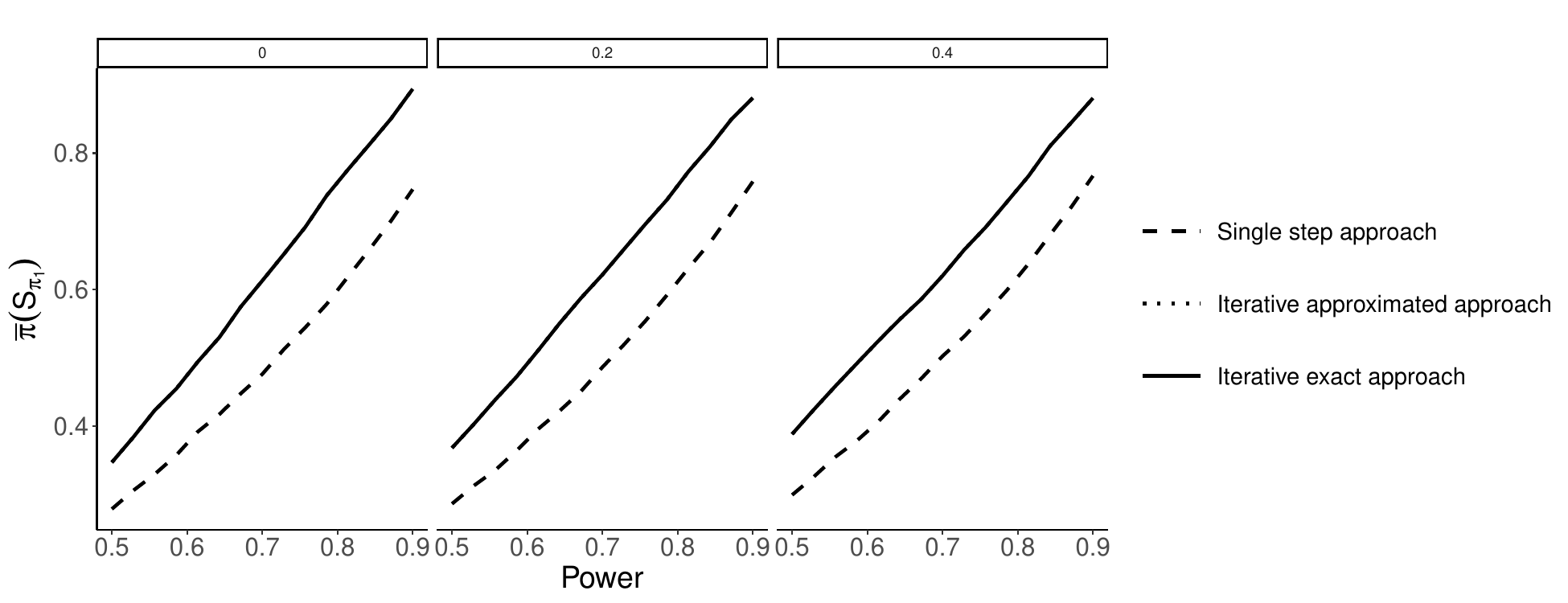}
	\caption{Simulated True Discovery lower bound for $S_{\pi_1}$ over different values of $\rho^2$ and power using the single-step (dashed line), iterative approximated version (dotted line) and iterative exact version (solid line). \angela{The dotted line is behind the solid line.}}
	\label{iterative_sim_equi}
\end{figure}

\angela{Then}, the performance of the iterative approach proposed in Section \ref{iterative} is compared with respect to the single-step one presented in Section \ref{methods}. Let consider directly $S_{\pi_1}$ the set of true discoveries, therefore $\pi(S_{\pi_1}) = 1$. Figure \ref{iterative_sim_equi} shows the true discovery proportion $\bar{\pi}(S_{\pi_1})$ computed on $1000$ simulated datas with $\pi_0$ equals $0.9$, $\rho^2 \in \{0, 0.2, 0.4\}$ and different levels of power used to simulate the data. In this case, we consider $m = 50$ so that we can use the exact iterative method. First of all, we can see how the approximated iterative version equals the exact one, and, more important, how both of them uniformly improve the single-step approach.

\begin{figure}
	\centering
	\includegraphics[width=.8\textwidth]{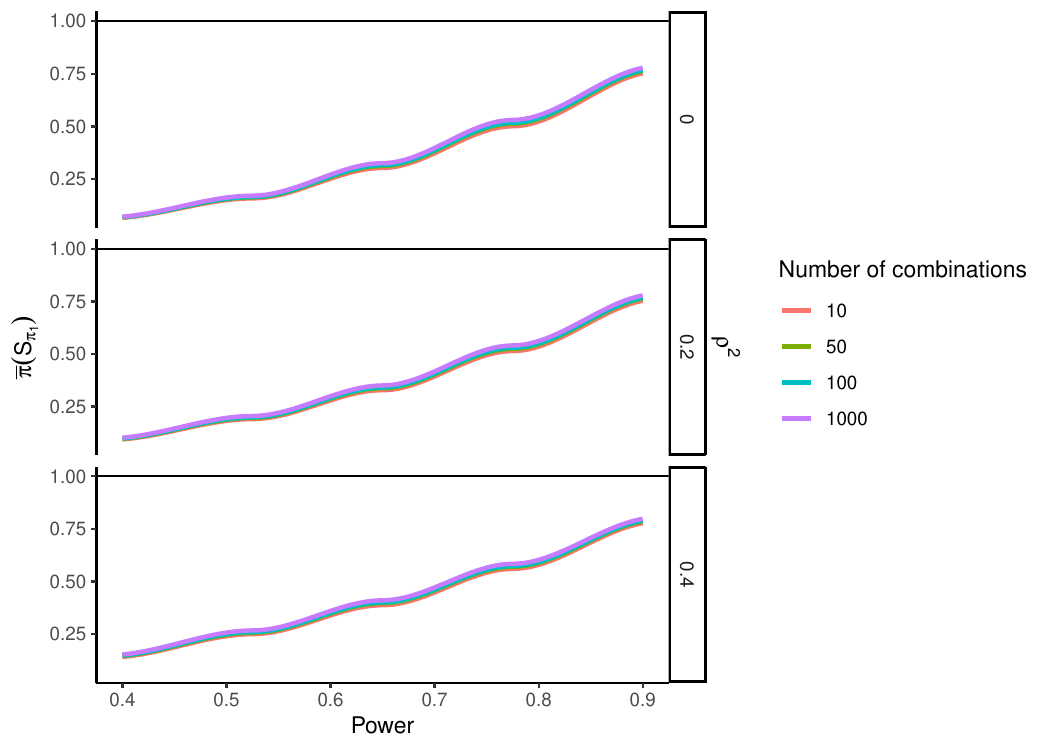}
	\caption{Simulated true discovery lower bounds for $S_{\pi_1}$ over different values of $\rho^2$ and power using the approximated iterative version with $10$, $50$, $100$, and, $1000$ random combinations. The solid black line represents the true discovery proportion $\pi(S_{\pi_1}) = 1$.}
	\label{fig:ncomb_equi}
\end{figure}

\begin{figure}
	\centering
	\includegraphics[width=.7\textwidth]{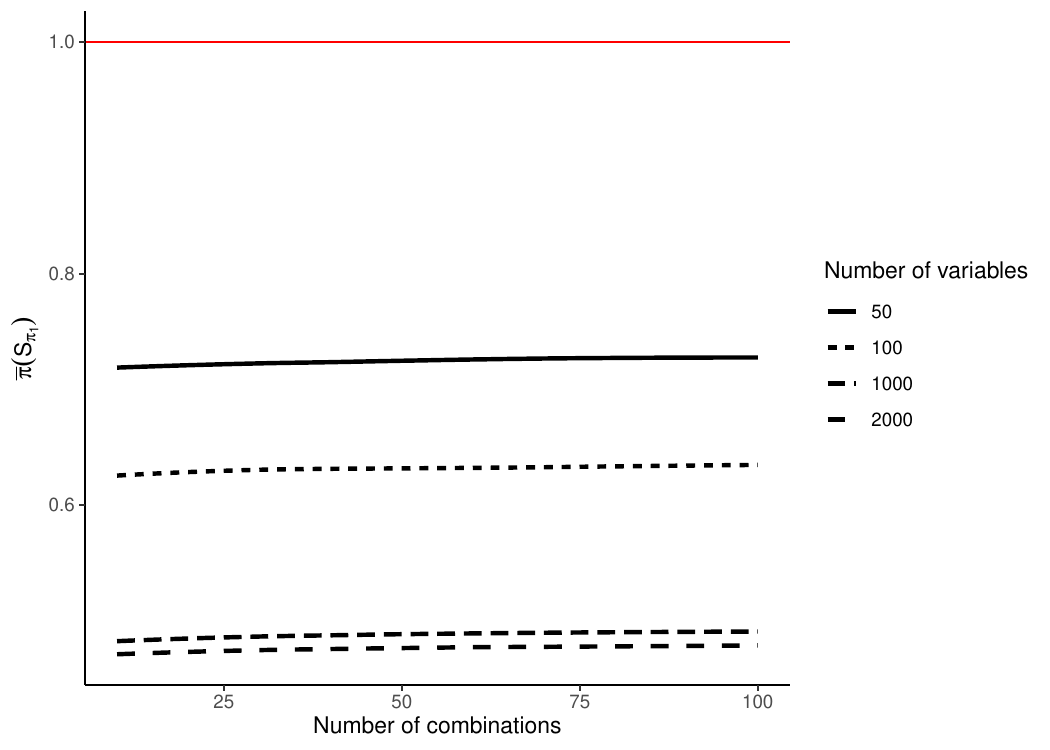}
	\caption{Simulated true discovery lower bounds for $S_{\pi_1}$ over different values of $m$ variables using the approximated iterative version with $10, \dots, 100$ random combinations. The solid red line represents the true discovery proportion $\pi(S_{\pi_1}) = 1$.}
	\label{fig:ncomb_m_equi}
\end{figure}

Figure \ref{fig:ncomb_equi} illustrates the behavior of the approximated iterative method using a different number of combinations. The method is applied directly on $S_{\pi_1}$ the set of true discoveries, therefore $\pi(S_{\pi_1}) = 1$. The approximation version becomes exact when the number of combinations goes to infinity. However, as we can see in Figure \ref{fig:ncomb_equi}, the results using only $10$ combinations are nearly equal to the results using $1000$ combinations. In addition, looking at Figure \ref{fig:ncomb_m_equi}, we can see that the method is robust if a different number of variables are considered, i.e., the lines in Figure \ref{fig:ncomb_m_equi} are below $1$ (true discovery proportion). 

To sum up, we suggest using the Higher Criticism and Beta families if the correlation across the variables is supposed to be low. Besides, we recommend considering the shifted version of the Simes or AORC family if the interest is in large sets of hypotheses rather than in small ones. and if the distribution of the p-value is expected to be anti-conservative, with a reasonable prior value of $\delta$ with respect to the data analyzed.

%
%
%

\newpage

\bibliographystyle{apalike}
\bibliography{bibliography.bib} 

\end{document}